\documentclass[12pt]{article}
\usepackage[reqno]{amsmath}
\usepackage{bbm}
\usepackage{epsfig}
\usepackage{array}
\usepackage{float}

\usepackage{a4}

\usepackage{a4wide}
\usepackage{wasysym}

\def\gs{\mathrel{
   \rlap{\raise 0.511ex \hbox{$>$}}{\lower 0.511ex \hbox{$\sim$}}}}
\def\ls{\mathrel{
   \rlap{\raise 0.511ex \hbox{$<$}}{\lower 0.511ex \hbox{$\sim$}}}}
\newcommand{\obb}{0\mbox{$\nu\beta\beta$}}
\newcommand{\onbb}{neutrinoless double beta decay }
\newcommand{\ba}{\begin{array}{c}}
\newcommand{\baz}{\begin{array}{cc}}
\newcommand{\bad}{\begin{array}{ccc}}
\newcommand{\bav}{\begin{array}{cccc}}
\newcommand{\bea}{\begin{equation} \begin{array}{c}}
\newcommand{\eea}{ \end{array} \end{equation}}
\newcommand{\ea}{\end{array}}
\newcommand{\D}{\displaystyle}
\newcommand{\dms}{\mbox{$\Delta m^2_{\odot}$}}
\newcommand{\dma}{\mbox{$\Delta m^2_{\rm A}$}}
\newcommand{\lsnd}{\mbox{$\Delta m^2_{\rm LSND}$}}
\newcommand{\meff}{\mbox{$\langle m \rangle$}}
\newcommand{\eV}{\mbox{ eV}}

\newcommand{\be}{\begin{eqnarray}}
\newcommand{\ee}{\end{eqnarray}}

\newcommand{\sss}{\sin^2 \theta_{\odot}}
\newcommand{\sch}{\sin^2 \theta_{\rm CHOOZ}}

\begin{document}

\title{\vspace{-2cm}
\hfill {\small TUM--HEP--615/05}\\
\vspace{-0.3cm}
\hfill {\small hep--ph/0512234} 
\vskip 0.5cm
\bf \large
Constraining Mass Spectra with Sterile Neutrinos from 
Neutrinoless Double Beta Decay, Tritium Beta Decay and Cosmology
}
\author{
Srubabati Goswami$^{a,b}$\thanks{email: \tt sruba@ph.tum.de}~~~and~~
Werner Rodejohann$^b$\thanks{email: \tt werner$\_$rodejohann@ph.tum.de} 
\\\\
$^a${\normalsize \it Harish--Chandra Research Institute, Chhatnag Road,}\\
{\normalsize \it Jhunsi, Allahabad 211 019, India}\\ \\
$^b${\normalsize \it Physik--Department, Technische Universit\"at M\"unchen,}\\
{\normalsize \it  James--Franck--Strasse, D--85748 Garching, Germany}
}

\date{}
\maketitle
\thispagestyle{empty}

\begin{abstract}

\noindent 
We analyze the  
constraints on neutrino mass spectra with extra sterile neutrinos 
as implied by the LSND experiment. 
The various mass related observables in 
neutrinoless double beta decay, tritium beta decay and 
cosmology are discussed. 
Both neutrino oscillation results as well as recent cosmological neutrino 
mass bounds are taken into account.
We find that some of the allowed mass patterns are severely restricted by the 
current constraints, in particular by the cosmological 
constraints on the total sum of neutrino masses and by 
the non-maximality of the solar neutrino mixing angle. 
Furthermore, we estimate the form of the four neutrino mass matrices and 
also comment on the situation in scenarios with two additional sterile 
neutrinos.

\end{abstract}
\newpage

\section{\label{sec:intro}Introduction}

Scenarios with four neutrinos became popular on the wake of the 
LSND evidence of $\bar{\nu}_\mu -\bar{\nu}_e$ transitions \cite{lsnd}. 
Interpreted in terms of neutrino oscillations, the indicated 
mass scale for LSND is in the eV$^2$ range.  
Together with the evidence for neutrino oscillations from 
atmospheric (plus K2K) and solar (plus KamLAND) neutrino observations, 
requiring mass scales around $10^{-3}$ eV$^2$ and 
$10^{-4}$ eV$^2$, respectively, 
a fourth sterile neutrino has to be introduced in order to 
accommodate the presence of three distinct mass squared 
differences.  

A priori, four neutrino scenarios allow for two 
possible mass patterns: 
\begin{itemize}
\item[(i)] 2+2 scenarios, in which two  
pairs of neutrino states are separated from each other by the 
LSND mass scale. 
There are two possibilities for 
2+2 scenarios; 
\item[(ii)] 3+1 scenarios, in which one single neutrino state is 
separated by the LSND mass scale from the other three states. 
There are four possibilities for 3+1 scenarios; 
\end{itemize}
Oscillation analyzes in both schemes were performed by a number of authors 
\cite{4mix1,4mix2,eitel,Giunti} and also the astrophysical and cosmological 
implications were investigated \cite{4cosmo,4cosmo_new}. 
Historically, among the above two alternatives 
the 3+1 scenarios were at first relatively disfavored \cite{4mix2} because 
of the non-observation of oscillations in short baseline experiments like 
KARMEN \cite{karmen}, Bugey \cite{bugey} and CDHS \cite{cdhsw}. Therefore 
the 2+2 scenarios were found to be more compatible with the existing data. 
The sterile neutrino oscillation solution in 2+2 scenarios was viable for 
both solar and atmospheric neutrinos. However, SuperKamiokande data 
disfavored oscillation of the atmospheric $\nu_\mu$ to purely sterile 
neutrinos \cite{SK_no_st}, and later on the SNO data 
started establishing the neutral current component 
in the solar $\nu_e$ flux \cite{sno}. 
For some time a mixed scenario, where 
the atmospheric neutrino anomaly is due to $\nu_\mu -\nu_{s,\tau}$ and 
the solar neutrino anomaly is due to $\nu_e - \nu_{s,\tau}$, 
remained compatible with all data \cite{sterilemix}. 
However, all recent analyzes show that 2+2 scenarios are ruled out
at a high $\sigma$ from the existing data \cite{vallesterile,valle_rev}. 
Both atmospheric and solar neutrino 
data strongly disfavor oscillations to pure sterile species. 
This disfavored the 2+2 scenarios irrespective of whether LSND results 
are confirmed or not. 
The most updated analysis in the 3+1 scheme performed in 
\cite{vallesterile,valle_rev} shows that non-evidence of neutrino 
oscillation in other short baseline (SBL) 
experiments combined with atmospheric neutrino data from SuperK and 
K2K is inconsistent with the LSND signal at 95\% C.L.\ and only marginal 
overlaps are found at 99\% C.L. 
Thus, with increased precision of solar and atmospheric neutrino flux 
measurements the four neutrino explanation of 
the LSND anomaly suffered a setback. 
This led to many alternative explanations of the LSND anomaly including 
introduction of two sterile neutrinos -- the so-called 3+2 scenario 
\cite{3plus2} --, CPT violation \cite{cpt}, 
quantum decoherence effects violating CPT \cite{decoh}, 
mass varying neutrinos \cite{mv}, 
neutrino decay in four neutrino scenarios \cite{4decay}, 
lepton number violating muon decay \cite{babu}, 
decay of a heavy neutrino \cite{sergio} or extra dimensional aspects 
\cite{extra}. 

Oscillation experiments can only measure the 
mass squared differences but not the absolute masses. 
The most direct and model independent way to measure 
the absolute masses is via 
kinematic measurements involving nuclear beta decay. 
The best bound at present is $m_{\beta} < 2.3$ eV (95\% C.L.) 
coming from the Mainz tritium beta decay experiment \cite{mainz}.  
The KATRIN experiment is expected to increase the sensitivity down to 
$\sim$ 0.2 eV \cite{katrin}.

Information on absolute masses can also come from 
neutrinoless double beta decay (\obb). 
Neutrinoless double beta decay experiments aim at 
observing the process 
\[ 
(A,Z) \rightarrow (A,Z + 2) + 2 \, e^-~. 
\] 
This is a lepton number violating process and its observation 
will establish the Majorana nature of neutrinos \cite{scheval}. 
The decay width depends quadratically on the so-called effective mass. 
We assume here that only the light Majorana neutrinos implied by 
neutrino oscillation experiments are exchanged in the diagram of \obb. 
In the basis in which the charged lepton mass matrix is 
real and diagonal, the 
effective mass is then nothing but the absolute value of the $ee$ element 
of the neutrino mass matrix. 
The best current limit on the effective mass is given by measurements of 
$^{76}$Ge established by the Heidelberg-Moscow collaboration \cite{HM} 
(with similar results obtained by the IGEX experiment \cite{IGEX}) 
\be \label{eq:current} 
\meff \le 0.35 \, \zeta~{\rm eV}~, 
\ee
where $\zeta={\cal O}(1)$ indicates that there is an uncertainty 
stemming from the nuclear physics involved in calculating the decay 
width of \obb.  The running projects NEMO3 \cite{NEMO} 
and CUORICINO \cite{CUORICINO} will be joined in the near future by 
next generation experiments such as CUORE \cite{CUORE}, 
MAJORANA \cite{MAJORANA}, GERDA \cite{GERDA}, EXO \cite{EXO}, 
MOON \cite{MOON}, COBRA \cite{COBRA}, XMASS, DCBA \cite{DCBA}, 
CANDLES \cite{CANDLES}, CAMEO \cite{CAMEO} (for a review see 
\cite{rev_ex}). One can safely expect that values of \meff{} 
one order of magnitude below the limit from Eq.\ (\ref{eq:current}) 
will be probed within the next, say, 
10 years\footnote{Not to forget, those experiments 
aim also to put the controversial \cite{contr} evidence of part of the 
Heidelberg-Moscow collaboration to the test.}. This means that  
scales of order $\sqrt{\lsnd}$ will be fully probed, 
and are even under investigation now. 
Since the effective mass measured in \obb\ also depends on the 
neutrino mixing angles, the neutrino mass scale and ordering, as well as 
the mass squared differences, 
it is possible to obtain additional constraints on sterile 
neutrino scenarios using neutrinoless double beta decay 
\cite{4others,4vbb,4others1}.

Important constraints on sterile neutrinos can 
also come from cosmology. 
Inclusion of an extra neutrino, even if sterile, can be in conflict with 
cosmological observations. The problems are increased if the extra sterile 
neutrino is massive and has significant mixing with the active species.  
In particular the Big Bang Nucleosynthesis model of standard cosmology, 
which explains light element abundances of the Universe, puts 
constraints on the number of neutrino species. 
The latest bound found in \cite{bbn1} for instance 
is $1.7 < N_{\nu} < 3.0$ at 95\% C.L.\  
and in \cite{bbn2} it is quoted that $N_{\nu}=3.14^{+0.70}_{-0.65}$. 
The differences in the results are due to different inputs regarding the
uncertainties in the primordial He abundance. 
Observations of the Cosmic Microwave Background and of 
large scale structures can also constrain the number of neutrino species. 
A summary of these bounds obtained by various groups including different data 
sets can be found in \cite{han05}. The upper limit on the number of neutrinos 
in these analysis can vary from 6 to 8. 
A recent bound as quoted in \cite{han05} is 
$N_{\nu} = 4.2 ^{+1.7}_{-1.2}$ at 95\% C.L.  
Another important constraint from cosmology comes on the sum of total 
masses of all the neutrinos, $\Sigma \equiv \sum m_i$. 
For four light neutrinos with degenerate masses the bound is 
$\Sigma < 1.7$ eV (95\% C.L.) from WMAP and 2dF data \cite{han05}. 
For four (five) neutrinos, with one (two) of them carrying a mass, 
the bound is $\Sigma < 1.05~(1.64)$ eV (95\% C.L.) \cite{hr0}. 
Improvement of these numbers within one order of magnitude is expected 
\cite{han05}. Note that these bounds 
depend on the priors and data sets used, for slightly more stringent 
bounds see, e.g., \cite{Dodelson:2005tp}. 
The above constraints can however be evaded if the abundances of 
sterile neutrinos in the early Universe can be suppressed. This requires   
going beyond the framework of standard cosmology and introducing mechanisms 
such as primordial lepton asymmetries \cite{foot}, low re-heating temperature
\cite{gelmini}, 
additional neutrino interactions \cite{bell}\footnote{See 
however \cite{hr}.} etc.  

Turning back to oscillations, the MiniBooNE experiment \cite{miniboone} 
is expected to confirm or refute the LSND signal and is expected to 
publish results within the next 6 months or so. 
If MiniBooNE does not confirm the LSND signal, then with the data collected 
with $10^{21}$ protons on target they can rule out the entire 
90$\%$ area allowed by LSND with 4 to 5$\sigma$ \cite{miniboone}. 
If however they confirm the LSND signal then this will give rise to an 
intriguing situation in what regards the explanation
of global oscillation data from accelerator, reactor, atmospheric and 
solar neutrino experiments. If confirming the LSND result, 
MiniBooNE can not distinguish 
the allowed four (or five) neutrino mass spectra. 
To understand the implied mass and mixing scheme, other 
observables are therefore crucial. This concerns in particular observables 
depending on the neutrino masses and ordering. Inasmuch one can 
use these future measurements to identify the neutrino spectrum 
is one of the motivations of this work. We stress here that we assume 
only the neutrino oscillation explanation of the LSND result is correct, 
i.e., the new physics alternatives (not necessarily predicting 
a signal for MiniBooNE) put forward in 
Refs.~\cite{cpt,decoh,mv,4decay,babu,sergio,extra} are not 
required. \\

In this paper we examine what constraints from current and future 
data can be obtained on possible neutrino mass spectra in scenarios 
with one or more sterile neutrinos. For the four allowed 
3+1 scenarios we give the neutrino masses, their sum as 
testable in cosmology, the kinematic neutrino mass for tritium experiments 
and the effective mass in \onbb$\!\!$. 
We include the most recent values of mass-squared 
differences and mixing angles from latest global analyzes 
of oscillation data. 
We furthermore reconstruct the possible mass matrices in four neutrino 
scenarios that are consistent with the current data. Finally, 
we also comment on 3+2 scenarios. 

The paper is build up as follows:  
In Section \ref{4nu} we discuss our parametrization of the 
four neutrino mixing matrix and summarize the relevant formulae for 
the neutrino masses, their sum, the kinematic neutrino 
mass measured in beta decay experiments and the effective mass 
that can be observed in neutrinoless double beta decay. 
In Section \ref{sec:0vbb3+1} we apply this framework to 3+1 scenarios. 
Approximate forms of the neutrino mass matrices in 3+1 schemes 
that are consistent with the current 
data are given in Section \ref{sec:matrices}. 
In Section \ref{sec:3+2} we comment on the above quantities 
in the 3+2 scheme, before presenting our summary and conclusions in 
Section \ref{sec:concl}. 
The oscillation probabilities for the relevant 
short baseline oscillation experiments are delegated to the Appendix. 
Although 2+2 scenarios are highly disfavored we also add for the sake 
of completeness an Appendix on the implications of such scenarios for 
neutrino masses from cosmology, beta decay and neutrinoless double beta decay. 
We also discuss the form of mass matrices in the 2+2 scenarios.  

\section{\label{4nu}Four Neutrino Mixing and Neutrino Masses} 

Neutrino mixing is described by the unitary 
Pontecorvo-Maki-Nakagawa-Sakata (PMNS) matrix $U$ \cite{PMNS}. For 
four Dirac neutrinos it contains 6 angles 
$\theta_{12,13,14,23,24,34}$ 
and three phases $\delta_{13, 14, 24}$ (``Dirac phases''). 
Their Majorana nature, which we shall assume, adds another three 
phases (``Majorana phases''), 
which do not have any consequences in neutrino oscillations \cite{MajPha}.
We parametrize $U$ as 
\bea \label{eq:Upara}
U = R_{34} \, \tilde R_{24} \,  \tilde R_{14} \, R_{23} \,  
\tilde R_{13} \, R_{12} \, P 
= 
\left( 
\bav 
U_{e1} & U_{e2} & U_{e3} & U_{e4} \\
U_{\mu 1} & U_{\mu 2} & U_{ \mu 3} & U_{ \mu 4}\\
U_{\tau 1} & U_{\tau 2} & U_{\tau 3} & U_{\tau 4}\\
U_{s 1} & U_{s 2} & U_{s 3} & U_{s 4}
\ea 
\right)
~, 
\eea
\begin{figure}[tb]
\vspace{1cm}
\hspace{-1cm}\epsfig{file=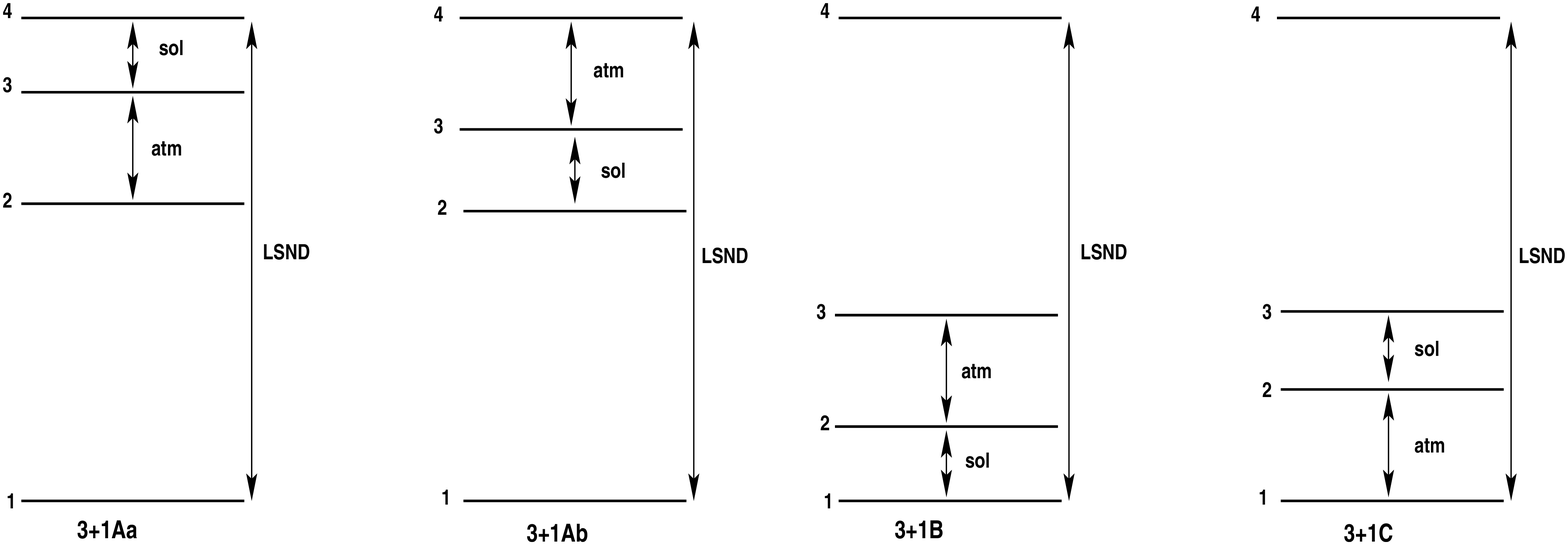,width=18cm,height=9.5cm}
\caption{\label{fig:3+1}The four allowed 3+1 mass orderings. }
\end{figure}
\begin{figure}[tb]
\hspace{.3cm}
\epsfig{file=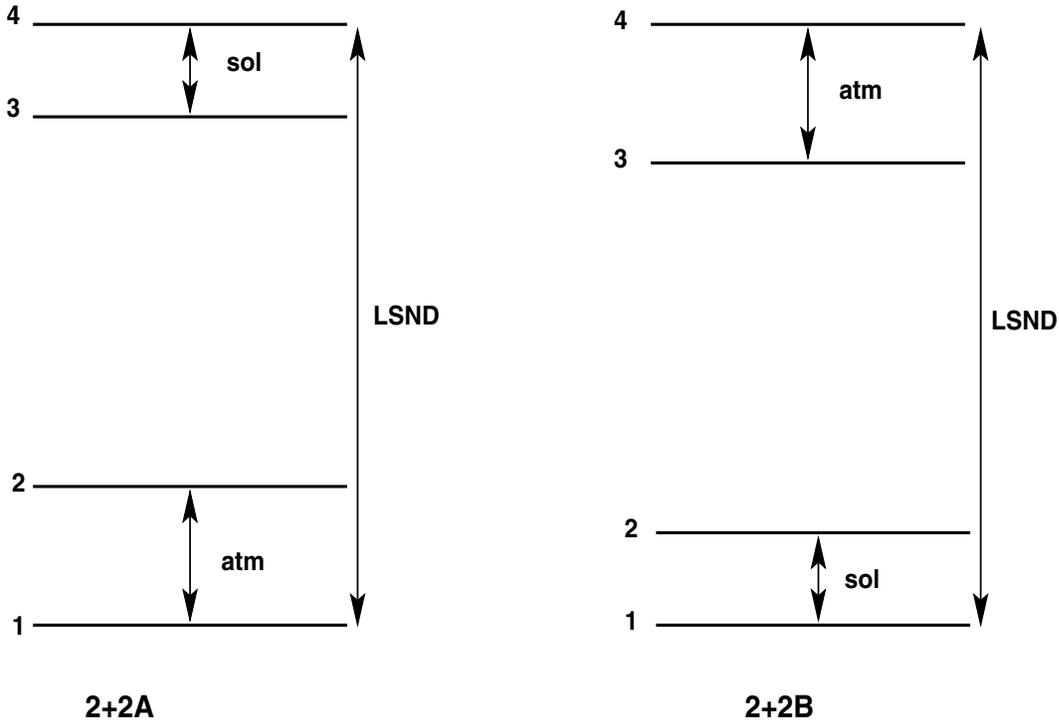,width=14cm,height=9.5cm}
\caption{\label{fig:2+2}The two allowed 2+2 mass orderings.}
\end{figure}
where the $R_{ij}$ represent rotations in $ij$ generation space, 
for instance:  
\be
R_{34} = 
\left( 
\bav 
1 & 0 & 0 & 0 \\ 
0 & 1 & 0 & 0 \\
0 & 0 & c_{34} & s_{34} \\ 
0 & 0 & -s_{34} & c_{34} 
\ea 
\right)~\mbox{ or }~ 
\tilde R_{14} = 
\left( 
\bav
c_{14} & 0 & 0 & s_{14} \, e^{-i \delta_{14}} \\ 
0 & 1 & 0 & 0 \\
0 & 0 & 1 & 0 \\ 
-s_{14} \, e^{i \delta_{14}} & 0 & 0 & c_{14} 
\ea 
\right)~,
\ee
with the usual notation $s_{ij} = \sin \theta_{ij}$ and 
$c_{ij} = \cos \theta_{ij}$. The diagonal matrix $P$ contains the three 
Majorana phases, which we denote $\alpha, \beta$ and $\gamma$: 
\be \label{eq:P}
P = {\rm diag} \left(1, e^{-i \alpha/2}, e^{-i(\beta/2 - \delta_{13})},
e^{-i(\gamma/2 - \delta_{14})} \right)~.
\ee
For most purposes it is sufficient to analyze the individual 
experimental data in a two-flavor framework. Depending on the neutrino 
mass spectrum, one can then identify certain elements of the PMNS 
matrix with the mixing angle in a two-neutrino oscillation probability. 
For the parameters governing solar (and KamLAND), atmospheric (and K2K) 
and short baseline reactor neutrino oscillation it holds at 3$\sigma$ 
\cite{sol_dat,lisiglobal}
\be
&7.0 \times 10^{-5} \eV^2 < \dms <  9.3\times 10^{-5} \eV^2~,&
\label{eq:msrange}
\\
& 0.25 < \sss < 0.40~, &
\label{eq:sssrange}
\ee
with best-fit values of $\dms = 8 \times 10^{-5}$ eV$^2$ and $\sss=0.31$. 
The atmospheric mass squared difference and mixing
angle at $3\sigma$ are known within \cite{valle_rev}
\be
&1.3 \times 10^{-3} \eV^2 < \dma < 4.2 \times 10^{-3} \eV^2&,
\label{eq:marange}
\\
&0.33 < \sin^2\theta_{\rm A} < 0.66,&
\ee
with best-fit values of $\dma = 2.2 \times 10^{-3}$ eV$^2$
and $\sin^2\theta_{\rm A}=1/2$.
The mixing angle $\theta_{13}$ at $3\sigma$ is restricted to lie
below the value \cite{chooz}
\be
\sch < 0.044~.
\label{eq:schrange}
\ee
In Appendix \ref{appB} we give the expressions for the oscillation 
probabilities in short baseline accelerator and 
reactor experiments and also for the 
1 km reactor experiment CHOOZ. 
For all the short baseline experiments excepting CHOOZ the  
one mass scale dominance approximation holds to a good precision. 
A comparison of the probabilities in the 3+1 picture (recall that 
2+2 scenarios are highly disfavored) with the 
two generation LSND probability reveals that 
$\sin^2 2 \theta_{\rm LSND}$ in 3+1 scenarios is always of the form 
$4 \, |U_{ei}|^2 \, |U_{\mu i}|^2$, where $|U_{ei}|^2$ is constrained 
to be small from Bugey reactor and solar neutrino data and 
$|U_{\mu i}|^2$ is constrained to be small from CDHS and atmospheric data. 
The index $i$ depends on the mass ordering. 
In Ref.\ \cite{vallesterile,valle_rev} the allowed area in the 
3+1 schemes is given in the 
$\Delta m^2_{\rm LSND} - \sin^2 2\theta_{\rm LSND}$ plane. 
The plot shows two overlap points at 99\% C.L.\  
corresponding to 
\be
(\Delta m^2_{\rm LSND}, \sin^2 2\theta_{\rm LSND}) = (0.9~{\rm eV}^2,0.002) 
\mbox{ and } (1.8~{\rm eV}^2,0.001)~.
\ee
In our following analysis we take these two allowed values 
as the illustrative values 
of $\Delta m^2_{\rm LSND}$ and take $|U_{ei}|^2$ as 0.008 and 0.004,  
respectively\footnote{This corresponds to $|U_{\mu i}|^2$ = 0.065, 
which is the highest allowed value of $|U_{\mu i}|^2$ 
at 99\% C.L.\ according to \cite{vallesterile}.}. 
It is to be noted that the MiniBooNE sensitivity plots in, e.g., 
Ref.\ \cite{sorel}, give an allowed region around $\lsnd = 0.9$ eV$^2$, 
which would be obtained if they confirm the LSND signal \cite{sorel}. 
Another allowed region is found for which 
\be
\Delta m^2_{\rm LSND} = (0.2 - 0.5)~{\rm eV}^2 \mbox{ and } 
\sin^2 2\theta_{\rm LSND} = 0.01 - 0.04~.
\ee
We therefore give in the following another set of plots for which we 
allow $\Delta m^2_{\rm LSND}$ and $\sin^2 2\theta_{\rm LSND}$ to vary in 
this range. 
To extract $|U_{ei}|^2$ from $\sin^2 2\theta_{\rm LSND}$
we assume that $|U_{ei}| \simeq |U_{\mu i}|$. 
With this approximation we have $4 \, |U_{ei}|^4 = \sin^2 2\theta_{\rm LSND}$, 
which gives $|U_{ei}|^2 \simeq 0.05 - 0.1$. 
It is to be noted that this range is not allowed at 
99\% C.L.\ according to the analysis of \cite{vallesterile}. 
Nevertheless we take this range as an illustrative example to 
compare with the other two cases with a relatively higher 
$\lsnd$ and lower $\sin^22\theta_{\rm LSND}$.\\

The neutrino mass matrix is given by 
\be \label{eq:mnu1}
m_\nu = U^\ast \, m_\nu^{\rm diag} \, U^\dagger~,\mbox{ where } 
 m_\nu^{\rm diag} = {\rm diag}(m_1, m_2, m_3, m_4)~.
\ee
We will order the four neutrinos such that $m_4 > m_3 > m_2 > m_1$ so that 
\lsnd{} is always given by $\lsnd = m_4^2 - m_1^4 \equiv \Delta m^2_{41}$. 
\begin{figure}[htb]
\begin{center}
\hspace{-2cm}
\epsfig{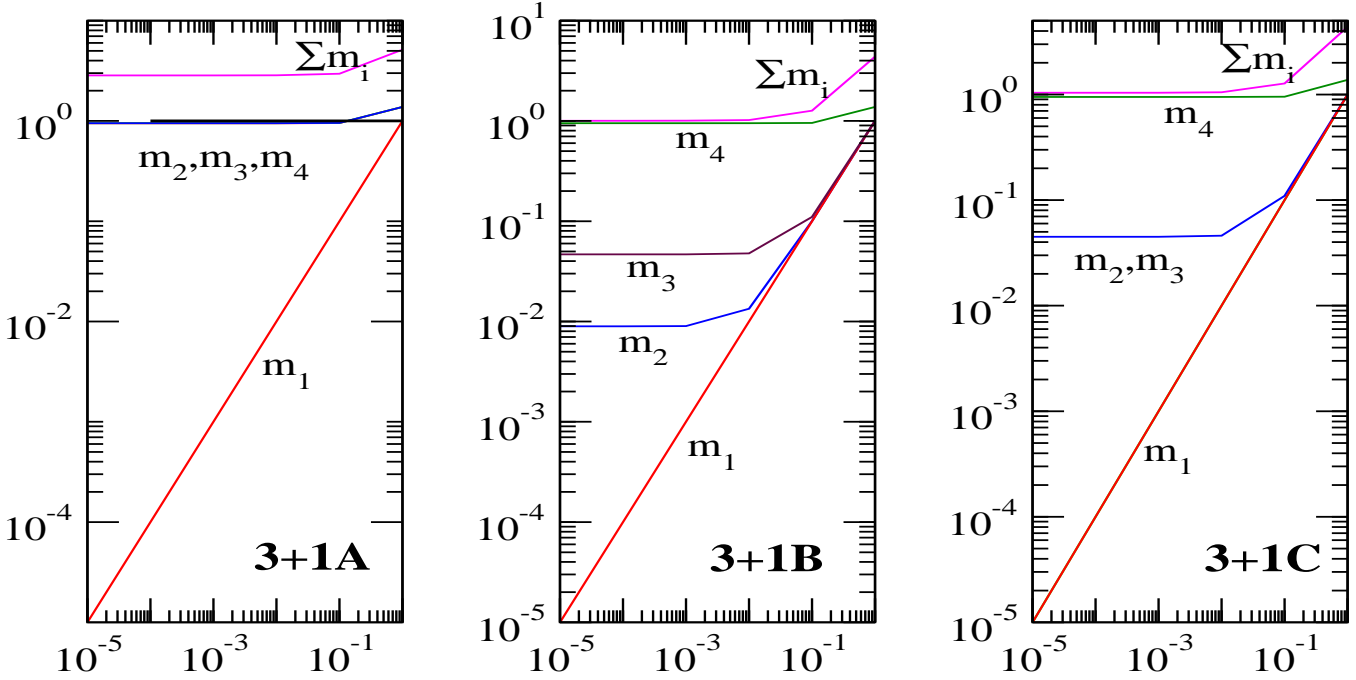}
\caption{\label{fig:3+1_mass}The individual neutrino masses $m_{1,2,3,4}$ 
and their sum $\Sigma$ as a function of the smallest neutrino mass $m_1$. 
We have 3+1Aa and 3+1Ab on the left (they generate basically 
identical results), 3+1B in the middle and 
3+1C on the right. We chose $\lsnd = 0.9$ eV$^2$ and fixed \dms\ and \dma\ 
to their best-fit values.}
\end{center}
\end{figure}
Depending on the relative ordering within the scheme, there are 
four possibilities for 3+1 and two possibilities for 2+2. 
We display all six cases in Figs.\ \ref{fig:3+1} and \ref{fig:2+2}. 
One might compare this situation with the three-flavor case, where 
there are only two possibilities, the normal and inverted ordering. 
The other two mass differences correspond to $\Delta m^2_{\odot}$ and 
$\Delta m^2_{\rm A}$. One can determine the individual masses as 
functions of the smallest mass $m_1$ and the three mass squared 
differences: 
\bea \label{eq:masses}
m_2 = \sqrt{m_1^2 + \Delta m^2_{21}} ~,\\[0.2cm]
m_3 = \sqrt{m_1^2 + \Delta m^2_{21} + \Delta m^2_{32}} ~,\\[0.2cm]
m_4 = \sqrt{m_1^2 + \Delta m^2_{21} + \Delta m^2_{32} + \Delta m^2_{43}} 
=  \sqrt{m_1^2 + \lsnd}~.
\eea
As an immediate application, we can then calculate the sum of 
neutrino masses $\Sigma$,
\be
\Sigma = m_1 + m_2 + m_3 + m_4~.
\label{cosmo}
\ee
for which interesting constraints from 
cosmology apply. Since the individual neutrino masses $m_{1,2,3,4}$ depend 
crucially on the mass spectrum, their sum $\Sigma$ will do so as well. 
For the four possible 3+1 neutrino spectra from Fig.\ \ref{fig:3+1} we 
display the neutrino masses and their sum $\Sigma$ as a function of the 
smallest mass in Fig.\ \ref{fig:3+1_mass}.

When in addition the mixing matrix elements of the PMNS matrix 
are specified, one can determine $m_{\beta}$, the parameter measured in 
the direct neutrino mass searches in nuclear beta decay 
experiments such as KATRIN. We will denote this parameter 
the ``kinematic mass''. 
It is given by: 
\be
m_{\beta} = \sqrt{|U_{e1}|^2 \, m_1^2 + |U_{e2}|^2 \, m_2^2 + 
|U_{e3}|^2 \, m_3^2 + |U_{e4}|^2 \, m_4^2 }~. 
\ee
With the inclusion of mixing, the parameters entering the 
mass measured in beta decay can be expressed in terms of 
the lowest mass, the mixing matrix elements $|U_{ei}|^2$ 
and the mass squared differences.
Consequently this quantity can 
also put constraints on the possible mass schemes and their ordering
\cite{farzan}. For three neutrino frameworks this has no observable 
effect since the future sensitivity on $m_\beta$ corresponds to 
quasi-degenerate neutrinos, for which unitarity of the PMNS matrix leads 
to no dependence on the mixing matrix elements and for which 
the normal and inverted ordering 
generate identical results. In the various four neutrino scenarios to 
be discussed in the following, this will change. 

Neutrinoless double beta decay experiments are 
sensitive to the effective mass which is given as 
\bea \label{eq:meff}
\meff = \left| \sum U_{ei}^2 \, m_i \right| = 
\left| |U_{e1}|^2 \, m_1 + |U_{e2}|^2 \, m_2 \, e^{i \alpha} 
+ |U_{e3}|^2 \, m_3 \, e^{i \beta} + 
|U_{e4}|^2 \, m_3 \, e^{i \gamma} \right|\\[0.3cm]
= 
\left| 
m_1 \, c_{12}^2 \, c_{13}^2 \, c_{14}^2 + m_2 \, e^{i \alpha} \, 
c_{13}^2 \, c_{14}^2 \, s_{12}^2 \, + m_3 \, e^{i \beta} \, 
c_{14}^2 \, s_{13}^2 + m_4 \, e^{i \gamma} \, s_{14}^2 \right| \\[0.4cm]
= c_{14}^2 \, \left| c_{13}^2 \, 
\left(m_1 \, c_{12}^2 + m_2 \, e^{i \alpha} s_{12}^2 
+ m_3 \, e^{i \beta} \, t_{13}^2 \right) +  m_4 \, e^{i \gamma} \, t_{14}^2 
\right|~. 
\eea
We defined here $t_{ij}=\tan \theta_{ij}$. As can be seen, \meff\ 
is sensitive to the Majorana phases which may 
be present in the neutrino mass matrix. The three Dirac phases 
do not appear in \meff. The effective mass depends on 10 out of the 16 
parameters of the general $4 \times 4$ neutrino mass matrix. 
This might be compared with the three-flavor case, in which \meff{} 
depends on 7 out of a total of 9 parameters. Moreover, as in the 
three-flavor case (for recent analyzes, see \cite{3meff,CR}), there 
is a strong dependence on the mass spectrum. 

We conclude that the three mass related observables \meff, $m_\beta$ and 
$\Sigma$ are powerful tools to discriminate among the 
various possible mass orderings. This will be the subject of the next 
Section.

\section{\label{sec:0vbb3+1}Neutrino Masses and 
Neutrinoless Double Beta Decay  
in 3+1 Scenarios} 
In the next Subsections we discuss the predictions 
for the sum of neutrino masses, 
the neutrino mass measured in nuclear beta decay experiments and the 
effective mass measured in neutrinoless double beta decay in the different 
3+1 scenarios. 
A common feature of the effective mass is that it can be expressed as 
a known three-flavor contribution obtained, e.g., in \cite{3meff,CR} 
plus an additional term related to the LSND scale. 

\subsection{\label{sec:0vbb3+1a}Neutrino Masses and 
Neutrinoless Double Beta Decay in Scenarios 3+1Aa and 3+1Ab}

As can be seen in Fig.\ \ref{fig:3+1},  
the scenarios 3+1Aa and 3+1Ab have three quasi-degenerate neutrinos 
with a mass given by the LSND scale and a fourth, lightest state 
separated by the LSND scale. 
It holds that $\dms = m_4^2 - m_3^2$, $\dma = m_3^2 - m_2^2$ and 
$\dms = m_3^2 - m_2^2$, $\dma = m_4^2 - m_3^2$, respectively. 
For the 3+1Aa case we can use Eq.\ (\ref{eq:masses}) to 
express the masses in terms of the smallest 
mass and the three mass squared differences as  
\bea
m_2 = \sqrt{m_1^2 + \lsnd - \Delta m^2_{\odot} -\dma}~,  \\[0.2cm]
m_3 = \sqrt{m_1^2 + \Delta m^2_{\rm LSND} - \Delta m^2_{\odot}} ~,\\[0.2cm]
m_4 = \sqrt{m_1^2 + \lsnd} ~.
\label{mass:3+1a}
\eea  
For the 3+1Ab case $\dma$ and $\dms$ replace each other. 
Since the neutrino masses are governed mainly by $\lsnd$, 
the predictions of scenarios 3+1Aa and 3+1Ab 
for all mass related observables are almost identical \cite{4others1} 
and therefore we treat these two cases together. 
Since the minimal mass of the 
three quasi-degenerate neutrinos is $\sqrt{\lsnd}$, 
it follows for the sum of neutrino masses 
that 
\be
\Sigma^{\rm 3+1Aa,b} \gs 3 \sqrt{\lsnd}~. 
\ee 
The left panel of Figure \ref{fig:3+1_mass} shows the 
four masses and their sum $\Sigma$ as a function of the smallest mass $m_1$. 
The three heavier masses are indistinguishable and much larger than 
the lightest state unless $m_1 \simeq 1$ eV. 
In Figure \ref{fig:3+1_mass} we have fixed $\Delta m^2_{\rm LSND}$ 
at 0.9 eV$^2$  and hence $\Sigma^{\rm 3+1Aa,b} \gs 2.8$ eV. 
It follows that these 
two schemes would already be in conflict with a cosmological limit 
of $\Sigma = 1$ eV unless $\lsnd \ls 0.1$ eV$^2$. 
Nevertheless, let us discuss these scenarios further. 
One would expect that $m_1 \ll m_{2,3,4}$ holds, so that  
the fourth state with mass $m_1$ effectively 
decouples in what regards the predictions. 

Neutrino data implies in scenario 3+1Aa (3+1Ab) that 
$\sin^2 2 \theta_{\rm LSND} = 4~|U_{e1}|^2 ~|U_{\mu 1}|^2$ 
(see Appendix \ref{appB}) and that 
$\sin^2 \theta_{\rm CHOOZ} \simeq |U_{e2}|^2$ 
($\sin^2 \theta_{\rm CHOOZ} \simeq |U_{e4}|^2$). 
One furthermore has $|U_{e3}|^2 \simeq \cos^2 \theta_\odot$ 
and $|U_{e4}|^2 \simeq \sin^2 \theta_\odot$ (case 3+1Aa) or 
$|U_{e3}|^2 \simeq \cos^2 \theta_\odot$ 
and $|U_{e2}|^2 \simeq \sin^2 \theta_\odot$ (case 3+1Ab). 

To an excellent approximation, the predictions for \onbb 
are very similar to the three-flavor case with 
quasi-degenerate neutrinos \cite{4others1}. 
In that case the normal and inverted mass ordering can not be 
distinguished via \obb, since their predictions differ only by 
corrections of order $\dma/m_0$, where $m_0$ is the common three-neutrino 
mass scale. 
Analogously, the scenarios 3+1Aa and 3+1Ab can not be distinguished via \obb, 
since they generate identical results up to corrections of order 
$\dma/\lsnd$. 
\begin{figure}[htb]
\centerline{\psfig{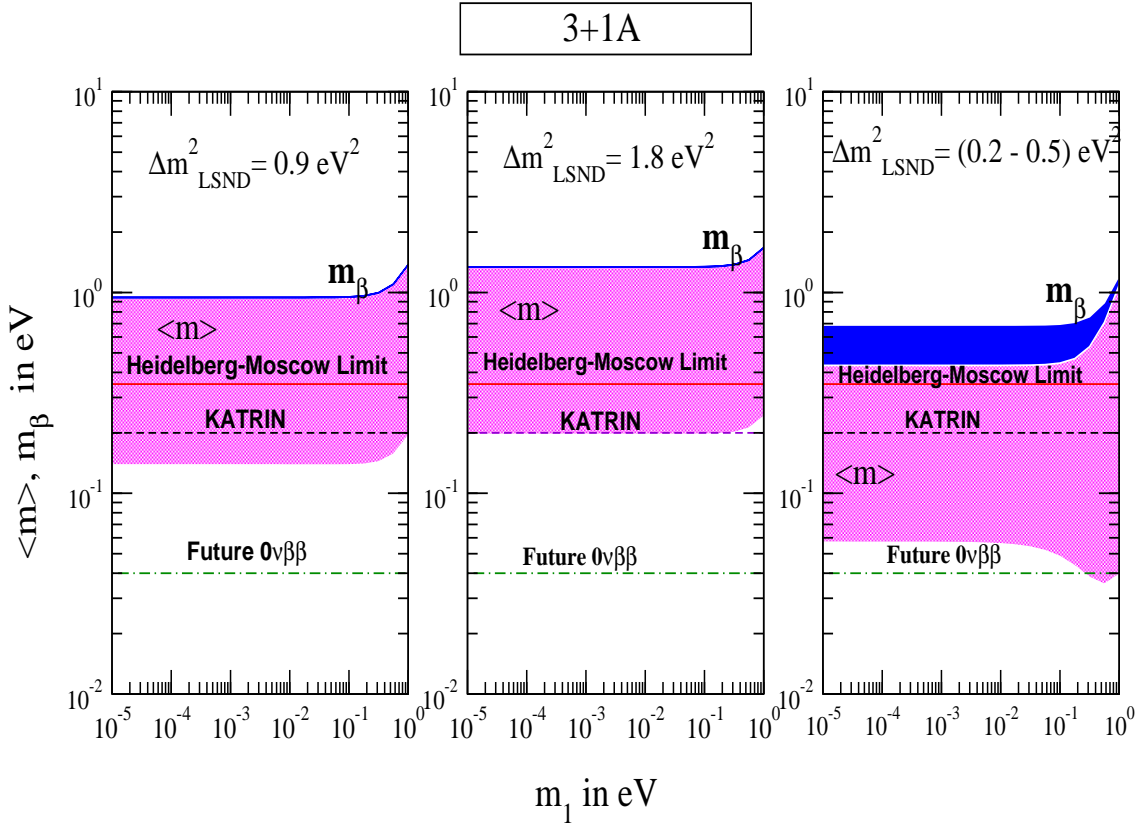}}
\caption{The effective mass as a function of the lowest mass in scenario 
3+1A. The left most column is for $\Delta m^2_{\rm LSND}$ = 0.9 eV$^2$ 
and the middle column is for $\Delta m^2_{\rm LSND}$ = 1.8 eV$^2$. 
The right most column is for $\Delta m^2_{\rm LSND}$ varied between 
(0.2 -- 0.5) eV$^2$. 
The other parameters are varied in their current 3$\sigma$ allowed range
and all the phases are varied between 0 and $2\pi$. 
Also shown is the mass $m_{\beta}$ that will be 
measured in beta decay experiments, the current and a prospective future 
limit on the effective mass and the future KATRIN limit.}
\label{fig:3+1a0vbb}
\end{figure}
For 3+1Aa, one has 
\bea
\meff^{\rm 3+1Aa} \simeq 
\left| 
\sqrt{\lsnd + m_1^2} \left( 
c_\odot^2 + s_\odot^2 \, e^{i(\alpha - \beta)} + \sin^2 \theta_{\rm CHOOZ} \, 
e^{i(\gamma - \beta)} \right) + e^{-i \beta} \, m_1 \, |U_{e1}|^2  
\right| \\[0.2cm]
\equiv \left| 
\meff^{\rm QD}_{3} +   m_1 \, e^{-i \beta} \, |U_{e1}|^2 
\right| ~.
\eea
We have defined $c_\odot = \cos \theta_\odot$, $s_\odot = \sin \theta_\odot$ 
and used that $\lsnd \gg \dma \gg \dms$. We also have defined 
\bea
\meff^{\rm QD}_{3} \equiv m_0~ \left(
\cos^2 \theta_\odot + e^{i(\alpha - \beta)} \sin^2 \theta_\odot + 
e^{i(\gamma - \beta)} \, \sin^2 \theta_{\rm CHOOZ} \right)\\[0.2cm]
\mbox{with } |\meff^{\rm QD}_{3}|
\ls m_0 
\frac{\D 1 - \tan^2 \theta_{12} - 2 \sin^2 \theta_{\rm CHOOZ}}
{\D 1 + \tan^2 \theta_\odot} \simeq m_0 \, \cos 2 \theta_\odot~,
\eea
with $m_0 = \sqrt{\lsnd + m_1^2}$. The quantity $\meff^{\rm QD}_{3}$ is 
just the usual three-flavor effective mass with the common mass scale 
given by the LSND scale. 
The contribution of 
$\meff^{\rm QD}_{3}$ is at least one order of magnitude above the 
term corresponding to the LSND scale: for instance, if $m_1=0.3$ eV, 
\lsnd = 0.9 eV$^2$, and with $\cos 2 \theta_\odot \simeq 0.2 - 0.5$, 
one has $\meff^{\rm QD}_{3} \gs 0.2$ eV and 
$m_1 \, \sin^2 2 \theta_{\rm LSND} \simeq 0.002$ eV. 

Hence, neglecting further $m_1^2$ 
with respect to \lsnd, and also $\sin^2 \theta_{\rm CHOOZ}$, one finds
\be
\meff^{\rm 3+1Aa} \simeq \sqrt{\lsnd} \sqrt{1 - \sin^2 2 \theta_\odot \, 
\sin^2 (\beta - \gamma)/2} ~.
\ee
Case 3+1Ab is obtained by replacing $\alpha$ with $\gamma$. 
Due to the non-maximal solar neutrino mixing, the effective mass 
can not vanish in case of scenarios 3+1Aa and 3+1Ab, its 
range is given by $\sqrt{\lsnd} \, \cos 2 \theta_\odot 
\ls \meff^{\rm 3+1Aa,b} \ls \sqrt{\lsnd}$. 
Therefore, for $\lsnd$ = 0.9 eV$^2$ it follows that 
$\meff_{\rm min} \simeq 0.19$ eV 
whereas for $\lsnd = 1.8$ eV$^2$ we have $\meff_{\rm min} \simeq 0.27$ eV.   
If $\lsnd$ is in the range (0.2 -- 0.5) eV$^2$, then $\meff_{\rm min}$ 
will be somewhat lower. 
This approximate behavior is reproduced in Figure \ref{fig:3+1a0vbb} 
where we have plotted the effective mass as a function of the 
smallest mass. Also included in the Figure is the predicted value of the 
parameter $m_\beta$, together with the anticipated KATRIN limit on 
$m_\beta$ of 0.2 eV. For comparison, we also gave the present 
Heidelberg-Moscow bound on the effective mass, together with a 
prospective future limit of 0.04 eV. 
Note that the minimum values of $\meff$ in the limit of 
vanishingly small $m_1$ are slightly lower than those mentioned above
because of the non-zero value of $\sin^2\theta_{\rm CHOOZ}$. 
The limit on the effective mass of $0.35 \, \zeta$ eV rules out 
part of the predicted range: for instance, if $\sin^2 \theta_\odot = 0.28$ 
and \lsnd = 0.9 eV$^2$, then $\sin^2 (\beta - \gamma)/2 \gs 0.93$ to 
obey the constraint of $\meff \ls 0.35$ eV. 
For $\sin^2 \theta_\odot = 0.31~(0.25; 0.40)$ and \lsnd = 0.9 eV$^2$, 
we have that $\meff \gs 0.36$ (0.48; 0.19) eV, whereas for 
\lsnd = 1.8 eV$^2$ it holds $\meff \gs 0.51$ (0.69; 0.27) eV. Therefore, 
if $\sin^2 \theta_\odot$ turns out to be on the lower side of its currently 
allowed range, then scenarios 3+1Aa,b face serious problems 
with the constraints from \obb. 

The kinematic neutrino mass in scenarios 3+1Aa,b is directly given by 
the LSND scale: 
\be
m_\beta^{\rm 3+1Aa,b} \simeq \sqrt{\lsnd} ~.
\ee
For fixed \lsnd\ the prediction for $m_\beta$ is therefore a line, whereas 
when a range of values for \lsnd\ is given, we also have a range 
of values for $m_\beta$. 

We can summarize the situation\footnote{We note here 
that the predictions for \obb{} and $m_\beta$ are 
identical 
(i.e., up to corrections of order $\dma/\lsnd$) to the  
one of (the highly disfavored) scenario 2+2A, treated in Appendix 
\ref{sec:appa}. Therefore, 
one can not distinguish these scenarios via \obb\ or tritium decay 
experiments. However, since one has $\Sigma^{\rm 2+2A} \simeq \frac 23 \, 
\Sigma^{\rm 3+1Aa,b}$, there would be a chance to distinguish them via 
cosmological measurements.} for scenarios 3+1Aa,b as follows: 
\be \label{eq:3+1A_sum}
\meff^{\rm 3+1Aa,b} \simeq m_\beta^{\rm 3+1Aa,b}\, 
\sqrt{1 - \sin^2 2 \theta_\odot \, \sin^2 \phi} \simeq 
\frac{\Sigma^{\rm 3+1Aa,b}}{3}\, 
\sqrt{1 - \sin^2 2 \theta_\odot \, \sin^2 \phi} ~,
\ee
where $\phi$ is some combination of Majorana phases. Some constraints 
on these parameters might be obtained in this scenario. 
The kinematic mass $m_\beta$ predicted by this scenario 
is much above the KATRIN sensitivity
and it will be disfavored if KATRIN confirms $m_\beta$ around 0.2 eV. 
The same is true 
when cosmological searches do not find a signal close to their 
current bounds. 
Future limits on the effective mass below roughly 0.05 eV will 
also rule out scenarios 3+1Aa,b. A limit of 0.1 eV will rule out the 
two overlap points at 0.9 and 1.8 eV$^2$.

\begin{figure}[htb]
\centerline{\psfig{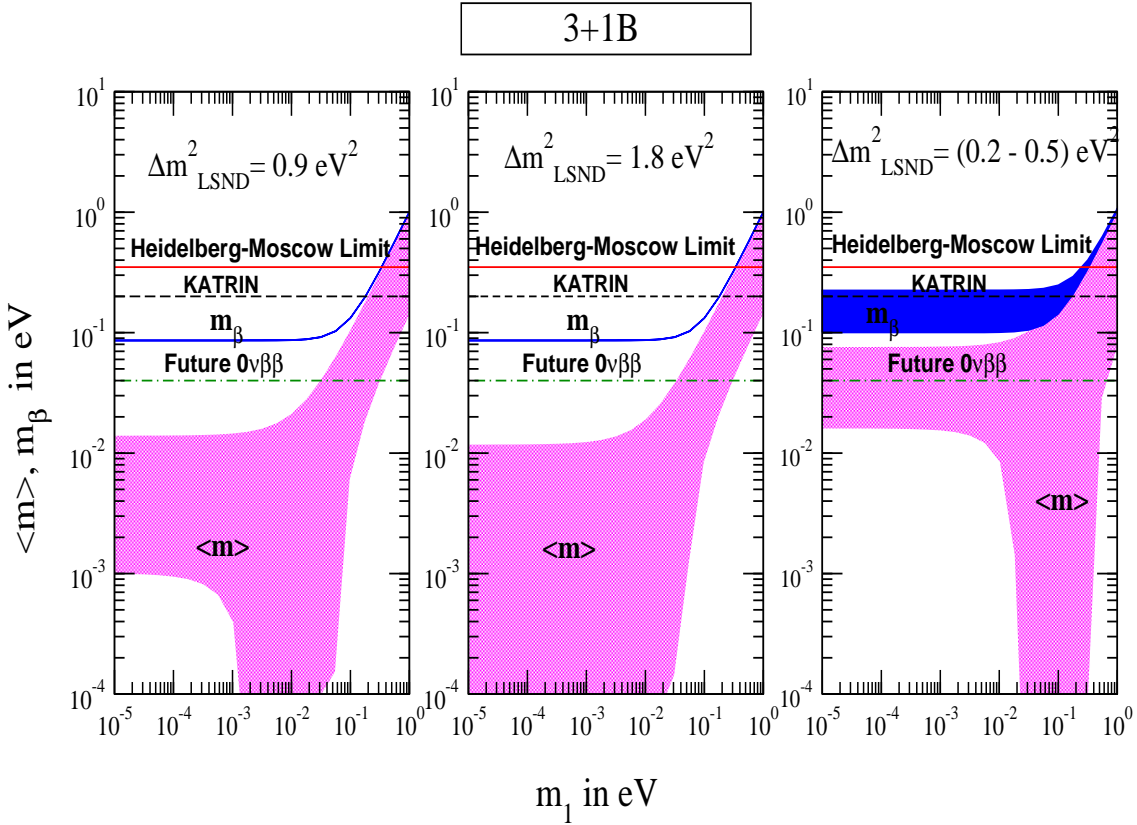}}
\caption{
Same as previous Figure for scenario 3+1B.}
\label{fig:3+1b0vbb}
\end{figure}

\subsection{\label{sec:0vbb3+1b}Neutrino Masses and Neutrinoless Double Beta 
Decay in Scenario 3+1B}

The structure of scenario 3+1B is depicted in Fig.\ \ref{fig:3+1}. 
We identify $\dms = m_2^2 - m_1^2$ and $\dma = m_3^2 - m_2^2$. 
The heaviest neutrino $m_4$ is separated by the LSND scale from the 
remaining three, which enjoy a normal ordering. 
We can express $m_2$, $m_3$ and $m_4$ 
in terms of the lowest mass $m_1$ and the three mass squared 
differences as   
\bea
m_2 = \sqrt{m_1^2 + \Delta m^2_{\odot} }~,  \\[0.2cm]
m_3 = \sqrt{m_1^2 + \Delta m^2_{\rm A} + \Delta m^2_{\odot} }~,\\[0.2cm]
m_4 = \sqrt{m_1^2 + \lsnd} ~.
\label{mass:3+1b}
\eea  
In Fig.\ \ref{fig:3+1_mass} we show -- with 
$\lsnd$ taken as 0.9 eV$^2$ -- the four masses 
as well as their sum as a function of the smallest mass $m_1$. 
We have ``unification'' of $m_2$ and $m_1$ when $m_1 \gs 0.01$ eV 
and of $m_3$ and $m_2$ when $m_1 \gs 0.1$ eV. 
For small $m_1 \ls 0.01$ eV one finds 
\be
\Sigma^{\rm 3+1B} \simeq \sqrt{\lsnd}~.
\ee 
To have $\Sigma \ls 1$ eV, it is required that $m_1 \ls 0.01~(0.1)$ eV if 
$\lsnd \simeq 1~(0.2)$ eV$^2$.

In this scheme it turns out that  
$|U_{e3}|^2 \simeq \sin^2\theta_{\rm CHOOZ}$ 
and $ 4~|U_{e4}|^2 ~|U_{\mu 4}|^2 = \sin^2 2\theta_{\rm LSND}$. 
The remaining elements of the $e$-row of $U$ are 
$|U_{e1}|^2 \simeq \cos^2 \theta_\odot$ and  
$|U_{e2}|^2 \simeq \sin^2 \theta_\odot$. 
Neglecting $m_1$, we have $m_2 \simeq \sqrt{\dms}$, 
$m_3 \simeq \sqrt{\dma}$ and $m_4 \simeq \sqrt{\lsnd}$. 
In this case, we can decompose the effective mass in a term well-known 
from three-flavor analyzes and a contribution from the LSND scale 
\cite{4others1}, namely: 
\bea
\meff^{\rm 3+1B} \simeq 
\left| \meff^{\rm NH}_3 + 
\sqrt{\lsnd} \, e^{i \gamma} \, |U_{e4}|^2 
\right|~,\mbox{ where } \\[0.2cm]
\meff^{\rm NH}_3 \equiv m_1 \, \cos^2 \theta_\odot + \sqrt{\dms}
 \, e^{i \alpha} \, 
\sin^2 \theta_\odot + \sqrt{\dma} \, e^{i \beta} 
\, \sin^2 \theta_{\rm CHOOZ}~.
\eea
The term  $|\meff^{\rm NH}_3|$ corresponds to the effective mass 
in case of three neutrinos with a normal hierarchy. 
The maximal value of $|\meff^{\rm NH}_3|$ in the range of $m_1^2 \ll \dms$ 
is known to be less than 0.007 eV \cite{3meff,CR}, whereas 
$ \sqrt{\lsnd} \, |U_{e4}|^2 \simeq 0.008$ eV  and 0.005 eV for the cases 
with $\lsnd$ = 0.9 eV$^2$ and 1.8 eV$^2$, respectively.  
Therefore, if $m_1$ is negligible and if 
$\lsnd$ = 0.9 eV$^2$, there can be no complete 
cancellation, and a lower (upper) limit on \meff\ of 
0.001 (0.015) eV can be expected. 
If $m_1$ is negligible and $\lsnd$ = 1.8 eV$^2$, however, 
there can be complete cancellation, and the upper limit of \meff\ is 
0.012 eV. For both values of \lsnd\ 
it turns out that also larger values for $m_1$ can lead to 
a vanishing effective mass, defining a ``cancellation regime'' in Fig.\ 
\ref{fig:3+1b0vbb}. 
If we vary $\lsnd$ in the range (0.2 -- 0.5) eV$^2$ and $m_1$ is small, then 
$\meff$ is controlled by the term $|U_{e4}|^2 \sqrt{\lsnd}$, 
which is between 0.02 and 0.07 eV, and consequently there is no complete 
cancellation. For higher values of $m_1$ there can be complete cancellation
in this case also. 
If the LSND scale is fixed to 0.9 or 1.8 eV$^2$ and $m_1 \gs 0.1$ eV, then 
the masses $m_{1,2,3}$ are quasi-degenerate and their contribution to 
the effective mass dominates the contribution from $m_4$. 
Consequently, the effective mass in this case is 
$\meff^{\rm 3+1B} \simeq m_0~(1 - \sin^2 2 \theta_\odot \, 
\sin^2 \alpha/2)$, 
where $m_0$ denotes the common mass scale of the three lightest 
neutrinos and $\theta_{\rm CHOOZ}$ has been neglected. 
For the MiniBooNE range of (0.2 -- 0.5) eV$^2$ this happens for 
slightly larger values of $m_1$, but the important aspect that 
the effective mass is non-zero in this case holds as well. 
The reason for this is that solar neutrino mixing is non-maximal. 
All the discussed features are reflected in Figure \ref{fig:3+1b0vbb}.

The kinematic neutrino mass is for $m_1=0$ 
\be
m_{\beta}^{\rm 3+1B} \simeq \sqrt{\dms \, \sin^2 \theta_\odot + 
\sin^2 \theta_{\rm CHOOZ} \, \dma  +  
|U_{e4}|^2  \, \lsnd }~, 
\ee
which is essentially determined by the term $\sqrt{\lsnd} \,  |U_{e4}|$. 
For $\lsnd = 0.9 $ eV$^2$ as well as $\lsnd = 1.8$ eV$^2$ this product is 
$\simeq 0.085$ eV, whereas for $\lsnd$ in the range (0.2 -- 0.5) eV$^2$ 
it can vary between (0.1 -- 0.2) eV. 
\begin{figure}[htb]
\centerline{\psfig{figure=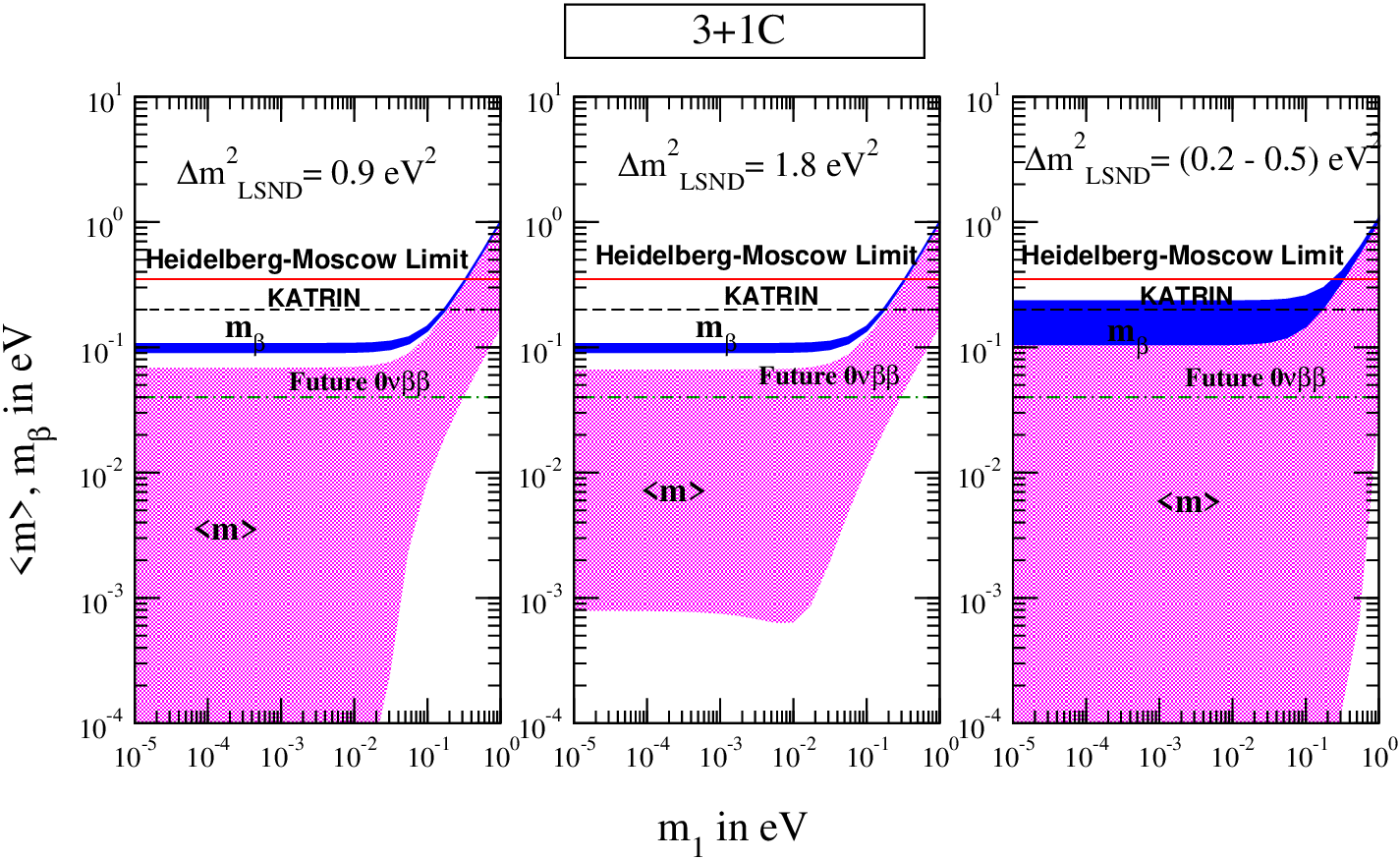,height=11cm,width=15cm}}
\caption{
Same as previous Figure for scenario 3+1C.} 
\label{fig:3+1c0vbb}
\end{figure}
This is reproduced in Figure \ref{fig:3+1b0vbb}. 
It is to be noted that $ m_{\beta}^{\rm 3+1B}$ 
can be one order of magnitude larger than the 
maximal effective mass. 
We can summarize scenario 3+1B for small $m_1$ as 
\be \label{eq:3+1b_sum}
\Sigma^{\rm 3+1B} \simeq \sqrt{\lsnd} > m_\beta^{\rm 3+1B} 
\gg \meff^{\rm 3+1B} ~.
\ee 
Looking at Figs.\ \ref{fig:3+1_mass} and \ref{fig:3+1b0vbb}, we 
can make the following statements:  
if cosmology improves the limit on $\Sigma$ to be below 
$\sqrt{\lsnd}$, scenario 3+1B can be ruled out. 
An effective mass above 0.05 (0.01) eV rules out both overlap points 
at $\lsnd = 0.9$ and 1.8 eV$^2$ when $m_1 \ls 0.1$ (0.01) eV is 
assumed to be small. 
In this case a successful KATRIN search will make this also possible.

\subsection{\label{sec:0vbb3+1c}Neutrino Masses and Neutrinoless Double Beta 
Decay in Scenario 3+1C}

The mass spectrum for the scenario 3+1C is depicted in Fig.\ \ref{fig:3+1}. 
We identify $\dms = m_3^2 - m_2^2$ and $\dma = m_3^2 - m_1^2$. 
The heaviest neutrino $m_4$ is separated by the LSND scale from the 
remaining three, which enjoy an inverted mass ordering. 
With this identification the different masses can be expressed in terms of 
the smallest mass $m_1$ and the mass squared differences as 
\bea
m_2 = \sqrt{m_1^2 - \Delta m^2_{\odot} + \Delta m^2_{\rm A} } ~,\\[0.2cm]
m_3 = \sqrt{m_1^2 + \Delta m^2_{\rm A}} ~,\\[0.2cm]
m_4 = \sqrt{m_1^2 + \lsnd }~.
\label{mass:3+1c}
\eea  
In Fig.\ \ref{fig:3+1_mass} we show the four masses 
as well as their sum as a function of the smallest mass $m_1$. 
We always have quasi-degeneracy between $m_2$ and $m_3$, and $m_1$ is 
quasi-degenerate with them once it is in the vicinity of 0.1 eV. 
The results on the total sum of masses 
are hardly distinguishable from 3+1B, 
\be
\Sigma^{\rm 3+1C} \simeq \sqrt{\lsnd}~.
\ee
For the mixing matrix elements one finds 
$|U_{e1}|^2 \simeq \sin^2\theta_{\rm CHOOZ}$ as well as 
$4\,|U_{e4}|^2 \, |U_{\mu 4}|^2 = \sin^2 2 \theta_{\rm LSND}$ and 
$|U_{e2}|^2 \simeq \cos^2 \theta_\odot$ and  
$|U_{e3}|^2 \simeq \sin^2 \theta_\odot$.

We 
show in Fig.\ \ref{fig:3+1c0vbb} our result for the effective mass as well 
as for the kinematic mass as a function of the smallest mass. 
The effective mass in the limit of small $m_1$ is 
\bea 
\meff^{\rm 3+1C} \simeq 
\left|
\sqrt{\dma} 
\left(e^{i \beta} \, \cos^2 \theta_\odot + e^{i \alpha} \, 
\sin^2 \theta_\odot\right)  + 
\sqrt{\lsnd} \, e^{i \gamma} \, |U_{e4}|^2\right|~,
\eea
which can also be written as  
\be
\meff^{\rm 3+1C} \simeq 
\left| \meff^{\rm IH}_3 + 
\sqrt{\lsnd} \, e^{i \gamma} \, |U_{e4}|^2 
\right| ~,
\ee
where the term  $|\meff^{\rm IH}_3|$ corresponds to the effective mass 
in case of three neutrinos with an inverted hierarchy \cite{4others1}. 
The contribution of the smallest mass $m_1$ plays a sub-leading role 
as it is also multiplied by small $\sin^2\theta_{\rm CHOOZ}$. 
The absolute value of $\meff^{\rm IH}_3$ is known to be 
$\sqrt{\dma} \sqrt{1 - \sin^2 2 \theta_\odot \, \sin^2 (\beta-\alpha)/2}$ 
and lies (for $m_1^2 \ll \dma$) between 0.007 and 0.05 eV \cite{3meff,CR}.
Whether there is complete cancellation in $\meff^{\rm 3+1C}$ 
or not depends on the value of $\sqrt{\lsnd} \, |U_{e4}|^2$. 
For $\Delta m^2_{\rm LSND} = 0.9~(1.8)$ eV$^2$ it is given by 
$\simeq 0.008~(0.005)$ eV. 
Therefore, for \lsnd = 0.9 eV$^2$ 
there can be complete cancellation as is seen in the first panel of 
Fig.\ \ref{fig:3+1c0vbb} but 
since for \lsnd = 1.8 eV$^2$ the value of $\sqrt{\lsnd} \, |U_{e4}|^2$ 
is less than 0.007 eV, there is no complete cancellation in this 
case (note that it is the other way around in scenario 3+1B). 
The lower limit on the effective mass is roughly $10^{-3}$ eV. 
Similarly, in the case where we vary \lsnd\ from (0.2 -- 0.5) eV$^2$ 
and $|U_{e4}|^2$ from 0.05 to 0.1, the value of the product  
$\sqrt{\lsnd} \, |U_{e4}|^2$  can be smaller than 0.06 eV and 
therefore in this case also there can be complete cancellation over a 
wide range of $m_1$. 
The maximum value of the effective mass in the small $m_1$ 
limit is roughly 
\bea
\meff^{\rm 3+1C}_{\rm max} \simeq 
\sqrt{\dma} +
\sqrt{\lsnd} \, |U_{e4}|^2~.
\eea
Since the product $\sqrt{\lsnd} \, |U_{e4}|^2$ is approximately the 
same for 0.9 and 1.8 eV$^2$, it follows that 
$\meff^{\rm 3+1C}_{\rm max}$ takes 
the same value ($\simeq 0.07$ eV) in these two cases. If $m_1$ is 
around 0.1 eV, then the three lightest masses are quasi-degenerate, and 
similar comments as for scenario 3+1B discussed in the previous Subsection 
apply. 

The kinematic neutrino mass for $m_1 \simeq 0$ is 
\be
m_{\beta}^{\rm 3+1C} \simeq \sqrt{\dma \,   
+ |U_{e4}|^2 \, \lsnd }~,
\ee
which is somewhat larger than the maximum effective mass. 
The blue (dark) band in Figure \ref{fig:3+1c0vbb} 
shows $m_{\beta}$ against $m_1$.
It is to be noted that in the first two columns of Figure 
\ref{fig:3+1c0vbb} the width of $m_{\beta}$ is due to the variation over  
the allowed range of $\dma$
as \lsnd\ is held fixed in these plots. 
In the third column the width is due to variation of both 
\lsnd\ and \dma. Let us summarize the situation: 
\be
\Sigma^{\rm 3+1C} \simeq \sqrt{\lsnd} \gg m_\beta^{\rm 3+1C} \gs 
\meff^{\rm 3+1C}_{\rm max}~.
\ee
Ruling out scenario 3+1C could be achieved if cosmology improves 
the limit on $\Sigma$ below 
$\sqrt{\lsnd}$. For small $m_1 \ls 0.1$ eV 
and a successful KATRIN search both overlap points 
at $\lsnd = 0.9$ and 1.8 eV$^2$ are ruled out. An effective mass above 
0.07 eV rules out all three cases under discussion, unless 
$m_1 \gs 0.1$ eV.

\section{\label{sec:matrices}Four Neutrino Mass Matrices} 

Models incorporating an extra sterile neutrino have been developed in many 
papers \cite{models,sruba,rabi_babu,roy_sudhir,lelmlt4}. 
In this Section we wish to summarize the typical mass matrices 
that are consistent with the experimental data in 3+1 scenarios. 
In particular, we look for simple $U(1)$ flavor symmetries 
which can force the approximate form of the mass matrices. 
The mass matrices for 
the 2+2 scenarios are discussed in Appendix \ref{sec:appa}. 

3+1 scenarios have the property that the sterile neutrino does
practically not participate in solar and atmospheric neutrino oscillations.
Consequently, there is very little dependence on the respective
sterile neutrino fraction.
Another general aspect of the results in 3+1 scenarios is that
the well-known three-flavor mass and mixing matrices (see the overviews in
\cite{3para,CR}) are ``embedded'' in the four-flavor
mass and mixing matrices. 
This means, in particular, that the corrections
to the usual three-flavor mass matrix are of order\footnote{We
have been made aware of an upcoming analysis on this
subject \cite{smirnov_talk}.}
$m_4 \, \sin^2 \theta_{\rm LSND} \sim \lambda^2 \, \sqrt{\lsnd}
\simeq \sqrt{\dms}$. We introduced here a small parameter 
$\lambda \simeq 0.1$, to estimate the different mass and mixing scales 
in the four neutrino framework. Both the LSND and the CHOOZ mixing angle 
are assumed to be of order $\lambda$, and the mass scales are related 
through $\dms \simeq \lambda^2 \, \dma \simeq 
\lambda^4 \, \lsnd$. 
In the approximation we are using, terms of order 
$\lambda^2 \, \sqrt{\lsnd}$ are subleading.

\subsection{\label{sec:mnu3+1a}The Mass Matrix in Scenarios 3+1Aa and
3+1Ab}
In case of scheme 3+1Aa, one has
\be
U^{\rm 3+1Aa} \simeq
\left(
\bav
\lambda & \lambda & \cos \theta_\odot & \sin \theta_\odot \\
\lambda & \sin \theta_A & -\sin \theta_\odot \, \cos \theta_A
& \cos \theta_\odot \, \cos \theta_A\\
\lambda & \cos \theta_A & \sin \theta_\odot \,  \sin \theta_A &
-\cos \theta_\odot \, \sin \theta_A \\
1 & \lambda & \lambda & \lambda
\ea
\right)\, P~,
\ee
where we set the Dirac phases to zero and included
small terms of order $\lambda \sim 0.1$
without writing possible order one coefficients. These terms indicate the
typical order of both the CHOOZ angle $\sin \theta_{\rm CHOOZ}$
and the LSND parameter $\sin \theta_{\rm LSND}$.
The above mixing matrix is unitary only to order $\lambda$.
In principle, the order one entries receive additional terms of order
$\lambda$ to cure this. The following analysis, however,
is not harmed by this.
With a given mass hierarchy we can obtain now the approximate form of the
mass matrix.
By looking at Fig.\ \ref{fig:3+1_mass}, we can see that
typically $m_4 \simeq m_3 \simeq m_2 \gg m_1$ holds.
So, setting $m_1=0$, we have
{\small
\be \nonumber\hspace{-.7cm}
m_\nu^{\rm 3+1Aa} \sim \sqrt{\lsnd} \,
\left(
\bav
e^{i \beta} \, c_\odot^2 + e^{i \gamma} \, s_\odot^2 &
c_{\rm A} c_\odot s_\odot (e^{i \beta} - e^{i \gamma}) &
s_{\rm A} c_\odot s_\odot (e^{i \beta} - e^{i \gamma})
& \lambda \\[0.2cm]
\cdot &  e^{i \alpha} \, s_{\rm A}^2  + c_{\rm A}^2
(e^{i \beta} \, s_\odot^2 + e^{i \gamma} \, c_\odot^2)
& c_{\rm A}  s_{\rm A}
(e^{i \alpha} - e^{i \gamma} c_\odot^2 - e^{i \beta} \, s_\odot^2)
& \lambda
\\[0.2cm]
\cdot & \cdot & e^{i \alpha} \, c_{\rm A}^2 + s_{\rm A}^2
(e^{i \beta} \, s_\odot^2 + e^{i \gamma} \, c_\odot^2) & \lambda \\[0.2cm]
\cdot & \cdot & \cdot & 0
\ea
\right)
\ee 
}
Here we have defined $c_{\rm A} = \cos \theta_A$ and 
$s_{\rm A} = \sin \theta_A$.
A matrix with
the entries of the $s$ column zero and the remaining elements of
order one conserves the flavor charge $L_s$.

\noindent
Obviously, the upper left $3 \times 3$-block of $m_\nu$ corresponds to the
well-known three flavor mass matrix in case of quasi-degenerate neutrinos.
Apart from the usual $\mu$-$\tau$ exchange symmetry \cite{mutau},
we can have several interesting special cases:
depending on the relative $CP$ parities of the three heavy neutrinos,
and setting for simplicity $\theta_{\rm A} = \pi/4$ and 
$\theta_\odot = \pi/4$, we can have $3 \times 3$
matrices proportional to the unit matrix ($\alpha = \beta = \gamma = 0$)
or with only a non-vanishing $ee$ and $\mu \tau$ element
($\beta = \gamma = \pi$ and $\alpha = 0$) \cite{3para}.\\
\noindent
The mixing matrix in scenario 3+1Ab is obtained by exchanging the second
and fourth row of the mixing matrix, i.e.,
\be
U^{\rm 3+1Ab} \simeq
\left(
\bav
\lambda & \sin \theta_\odot & \cos \theta_\odot & \lambda  \\
\lambda & \cos \theta_\odot \, \cos \theta_A
& -\sin \theta_\odot \, \cos \theta_A
& \sin \theta_A \\
\lambda & -\cos \theta_\odot \, \sin \theta_A
& \sin \theta_\odot \,  \sin \theta_A &
\cos \theta_A \\
1 & \lambda & \lambda & \lambda
\ea
\right)\, P~.
\ee
The mass matrix looks identical to case 3+1Aa, the only change being the
replacement $\alpha \leftrightarrow \gamma$.
This is analogous to the three-flavor case, in which the structure of
the mass matrix for quasi-degenerate neutrinos does not depend on
whether the neutrinos are normally and inversely ordered.

\subsection{\label{sec:mnu3+1b}The Mass Matrix in Scenario 3+1B}
The mixing matrix is given by
\be
U^{\rm 3+1B} \simeq
\left(
\bav
\cos \theta_\odot & \sin \theta_\odot
& \lambda & \lambda  \\
-\sin \theta_\odot \, \cos \theta_A &
\cos \theta_\odot \, \cos \theta_A
& \sin \theta_A  & \lambda \\
\sin \theta_\odot \,  \sin \theta_A &
-\cos \theta_\odot \, \sin \theta_A
& \cos \theta_A  & \lambda \\
\lambda & \lambda & \lambda & 1
\ea
\right)\, P  ~.
\ee
Regarding the mass states, we have for a smallest mass $m_1 \ls 0.005$ eV
that $m_4 \simeq \sqrt{\lsnd} \gg m_3 \simeq \sqrt{\dma} \simeq
m_4 \, \lambda \gg m_2 \simeq \sqrt{\dma} \simeq m_4 \, \lambda^2$.
The mass matrix then reads
\be \nonumber\hspace{-.7cm}
m_\nu^{\rm 3+1B} \sim \sqrt{\lsnd} \,
\left(
\bav
0 & 0 & 0 & \lambda \\[0.2cm]
\cdot & s_{\rm A}^2 \, \lambda & c_{\rm A}  s_{\rm A} \, \lambda &
\lambda \\[0.2cm]
\cdot & \cdot & c_{\rm A}^2 \, \lambda & \lambda \\[0.2cm]
\cdot & \cdot & \cdot & 1
\ea
\right)~.
\ee
Again, the upper left $3 \times 3$-block corresponds to the well-known
three-flavor mass matrix, which can be obtained by demanding $L_e$ to
be conserved\footnote{Conservation of $L_e$ in case of four flavors
would be achieved when the $\mu s$ and $\tau s$ entries are also of order
one.}.
A four-flavor mass matrix with only the $ss$ element
non-zero conserves\footnote{Obviously, conserving
$L_e + L_\mu + L_\tau$ in case of three flavors
leads to Dirac neutrinos. For four flavor scenarios one would have
to ask for $L_e + L_\mu + L_\tau + L_s$ conservation.}
$L_e + L_\mu + L_\tau$. As long as $m_1$ (and therefore also $m_2$ and $m_3$)
are one order of magnitude below $m_4$, the main structure of the mass
matrix remains, i.e., the $ss$ entry is one order of magnitude larger
than the remaining ones.

\subsection{\label{sec:mnu3+1c}The Mass Matrix in Scenario 3+1C}
The mixing matrix is given by
\be
U^{\rm 3+1C} \simeq
\left(
\bav
\lambda & \cos \theta_\odot & \sin \theta_\odot
& \lambda  \\
 \sin \theta_A  & -\sin \theta_\odot \, \cos \theta_A &
\cos \theta_\odot \, \cos \theta_A
& \lambda \\
 \cos \theta_A  & \sin \theta_\odot \,  \sin \theta_A &
-\cos \theta_\odot \, \sin \theta_A
&\lambda \\
\lambda & \lambda & \lambda & 1
\ea
\right)\, P  ~.
\ee
The three light neutrinos correspond approximately to the
well-known inverted hierarchy case of three neutrinos.
With $m_1 \simeq 0$ and $m_3 \simeq m_2 \simeq \sqrt{\dma} \simeq
m_4 \, \lambda \simeq \sqrt{\lsnd} \, \lambda$, we get
{\small
\be \nonumber
m_\nu^{\rm 3+1C} \sim \sqrt{\lsnd} \,
\left(
\bav
e^{i \alpha} c_\odot^2 + e^{i \beta} \, s_\odot^2 \, \lambda &
(e^{i \alpha} - e^{i \beta}) c_{\rm A} s_\odot  c_\odot \, \lambda &
(e^{i \alpha} - e^{i \beta}) s_{\rm A} s_\odot  c_\odot \, \lambda &
\lambda \\[0.2cm]
\cdot &
(e^{i \alpha} s_\odot^2 + e^{i \beta} \, c_\odot^2)  c_{\rm A}^2
\, \lambda &
(e^{i \alpha} s_\odot^2 + e^{i \beta} \, c_\odot^2)  c_{\rm A} s_{\rm A}
\, \lambda & \lambda \\[0.2cm]
\cdot & \cdot & (e^{i \alpha} s_\odot^2 + e^{i \beta} \, c_\odot^2)
s_{\rm A}^2 \, \lambda & \lambda \\[0.2cm]
\cdot & \cdot & \cdot & 1
\ea
\right)~.
\ee }
The full four-flavor matrix conserves approximately (i.e., 
when we neglect $\lambda$) the flavor charge $L_e + L_\mu + L_\tau$.
We have again for the upper left $3\times3$ block the well-known
three flavor mass matrix of an inverted hierarchy, which displays for
$\theta_{\rm A} = \pi/4$ a $\mu$-$\tau$ symmetry.

\noindent
If we set $\theta_\odot = \pi/4$ and choose\footnote{Note that 
such a Pseudo-Dirac structure will lead to enhanced stability 
with respect to radiative corrections.} $\alpha = 0$ and
$\beta=\pi$, then all entries except the $ss$, $e\mu$ and $e\tau$ elements
vanish to order $\lambda$:
\be
m_\nu^{\rm 3+1C} \sim \sqrt{\lsnd} \,
\left(
\bav
0 & 1 & 1 & 0 \\[0.2cm]
\cdot & 0 & 0 & 0 \\[0.2cm]
\cdot & \cdot & 0 & 0 \\[0.2cm]
\cdot & \cdot & \cdot & 1
\ea
\right)~.
\ee
The global symmetry forcing this form of the mass matrix is
$L_e - L_\mu - L_\tau$, which was introduced first for the three-flavor case
\cite{stp}, but was used also for the four flavor case \cite{lelmlt4}.

\section{\label{sec:3+2}Comments on 3+2 Scenarios}
By means of introducing more sterile neutrinos, the goodness of fit
for explaining the LSND and other short-baseline data
can, not really surprisingly, be improved.
In this respect, the 3+2 scenario has been put forward to make the
interpretation of all neutrino data less problematic \cite{3plus2}.
These schemes have three neutrinos actively oscillating among themselves
and {\it two} additional sterile neutrinos responsible for the 
LSND anomaly. 
Models to accommodate 3+2 scenarios can be found in \cite{32models}.
With 5 neutrinos participating in neutrino oscillations, 4 independent
$\Delta m^2$ are present. In addition to the three discussed previously, 
we have to deal in addition with $\Delta m^2_{51}$.

In the analysis of Ref.\ \cite{3plus2} two best-fit points are given,
one of which corresponds to $\Delta m^2_{51} = 22 \eV^2$,
which 
is clearly not consistent with  cosmological constraints. 
The second best-fit value is at 
\bea
\Delta m^2_{41} = 0.46 \eV^2~,~~ \Delta m^2_{51} = 0.89 \eV^2 ~,\\[0.2cm]
U_{e4}=0.090~,~~ U_{e5}=0.125~,~~ U_{\mu 4}=0.226~,~~  U_{\mu 5}=0.160 ~.
\eea
This identification of the mixing matrix elements assumes that the
three active neutrinos are lighter than the two sterile ones,
i.e., a situation resembling scenarios 3+1B and 3+1C. 
We can decompose the effective mass as a term from the three 
active neutrinos and a term from the two sterile ones, i.e., 
\be
\meff^{3+2} = 
\left| 
\meff^{\rm 3 \, ac} + \meff^{\rm 2 \, st} 
\right|~.
\ee
In this case, the two additional mass scales imply an additional
contribution to the effective mass, reading
\be
\meff^{\rm 2 \, st} \equiv
|U_{e4}|^2 \, m_4 \, e^{i \gamma} + |U_{e5}|^2 \, m_5 \, e^{i \rho} =
|U_{e4}|^2 \,  \sqrt{m_1^2 + \Delta m^2_{41}} \, e^{i \gamma} +
|U_{e5}|^2 \, \sqrt{m_1^2 + \Delta m^2_{51}} \, e^{i \rho}~,
\ee
where we have introduced a fourth relevant Majorana phase $\rho$.
If $m_1 = 0$, then the best-fit values given above yield
$|\meff^{\rm 2 \, st}| \simeq$ (0.01 -- 0.02) eV, where the range
is caused by the effect of the Majorana phases.

The cosmological mass parameter is $\Sigma \simeq
\sqrt{m_1^2 + \Delta m^2_{41}} + \sqrt{m_1^2 + \Delta m^2_{51}}$,
which for $m_1=0$ is 1.6 eV, remarkably close to the current 
relevant limit from cosmology (1.64 eV) obtained in Ref.\ \cite{hr}.

The contribution to the neutrino mass measurable in KATRIN
is roughly 
\bea
m_{\beta} \simeq \sqrt{|U_{e4}|^2 \, (m_1^2 + \Delta m^2_{41})
+ |U_{e5}|^2 \, (m_1^2 + \Delta m^2_{51}) }~
\eea
which for $m_1=0$ is 0.1 eV,
but can reach testable values if $m_1 \gs 0.1$ eV.

If the three active neutrinos display an inverted hierarchy, then
their contribution $|\meff^{\rm 3 \, ac}|$ 
to the effective mass (see scenario 3+1C in
Section \ref{sec:0vbb3+1c}) lies between 
$\meff = \sqrt{\dma}$ and $\meff = \sqrt{\dma} \, \cos 2 \theta_{\odot} $, or 
numerically: (0.007--0.06) eV.  
Complete cancellation between $\meff^{\rm 3 \, ac}$ and
$\meff^{\rm 2 \, st}$ is possible. On the other hand, positive 
interference of the two terms can lead to $\meff \simeq 0.08$ eV. 
If the three active neutrinos enjoy a normal mass ordering with $m_1=0$,
then their contribution $\meff^{\rm 3 \, ac}$ to \meff{}
does not exceed 0.007 eV (see scenario 3+1B in Section
\ref{sec:0vbb3+1b}), so that $\meff^{\rm 2 \, st}$ dominates and
represents basically the prediction for \meff. 
Note however that these considerations take use of the best-fit values
of the fourth and fifth neutrino sector. Varying these parameters within
their allowed range and/or effects of non-zero $m_1$
might easily allow for severe cancellation. Moreover, one might also
exchange $\Delta m^2_{41}$ and $\Delta m^2_{51}$, leading to 
identical values for $\Sigma$, slightly larger values of $m_\beta$ and 
$|\meff^{\rm 2 \, st}| \simeq (0.003 - 0.02)$ eV  (all if $m_1=0$). 
Consequently, the effective mass vanishes for both orderings 
of the three active neutrinos. The upper limit is 0.02 (0.08) eV 
for normally (inversely) ordered active neutrinos. 

Up to now the two sterile neutrinos were assumed be heavier than
the active ones. 
Also possible is that the two additional neutrinos are lighter than the
three active ones, thereby resembling scenarios 3+1Aa and 3+1Ab. 
For the mixing matrix elements one has to exchange the indices
$4 \leftrightarrow 1$ and $5 \leftrightarrow 2$.
The three active ones generate an effective mass larger than
roughly $\sqrt{\Delta m^2_{51} + m_1^2} \cos 2 \theta_\odot$, which for
$m_1 = 0$ is 0.1 eV (see Section \ref{sec:0vbb3+1a}).
The contribution of the additional neutrinos is significantly
suppressed. The KATRIN parameter is approximately
$\sqrt{\Delta m^2_{51}} \simeq 0.9$ eV, surely a testable value.
Cosmology will
have to probe $\Sigma \simeq 3 \sqrt{\Delta m^2_{51}} \simeq 2.8$ eV,
which is incompatible already with current limits.\\

Finally, it should be clear that the known three-flavor mass and 
mixing matrices are embedded in the $5\times5$ mass and mixing 
matrices. We note that scenarios with 3 or more sterile neutrinos will 
also show this ``embedding''.

\section{\label{sec:concl}Conclusions and Summary}

We  have examined the constraints on LSND induced scenarios with 
extra sterile neutrinos from current and future bounds 
on \onbb$\!\!$, tritium beta decays and cosmological limits on the 
sum of neutrino masses. 
Since 2+2 scenarios are already disfavored by the present 
solar and atmospheric data we considered the 3+1 scenario in the 
main part of the paper.
The values of $\lsnd$ considered in our analysis are 0.9 eV$^2$ and 
1.8 eV$^2$, allowed by a combined analysis of SBL + atmospheric and 
K2K data \cite{vallesterile,valle_rev}.
We also considered $\lsnd$ to vary in the range (0.2 -- 0.5) eV$^2$, 
motivated by MiniBooNE sensitivity plots. 
For sake of completeness we discuss the 2+2 scenarios in Appendix 
\ref{sec:appa}. 
Within the 3+1 scenario there are three possibilities, 3+1Aa,b, 3+1B and 3+1C. 
The sum of neutrino masses is 
$\simeq 3(1,1) \sqrt{\lsnd}$, for scenario 3+1Aa,b (3+1B, 3+1C) 
neglecting the solar and atmospheric mass differences.  
The effective mass can be written as a known three-flavor contribution plus 
an additional term stemming from the LSND scale. Let us summarize the 
different aspects: 
\begin{itemize} 
\item 
in the 3+1Aa and 3+1Ab scenarios we have three quasi-degenerate  
neutrinos of mass $\sqrt{\lsnd}$ and a fourth state separated from 
them by the LSND mass scale. 
The sum of the masses is $\simeq 3 \sqrt{\lsnd}$ in the limit where 
the mass of the lowest state is vanishingly small. 
This implies small values of $\lsnd$ ($\simeq 0.1$ eV$^2$)
to comply with the cosmological bound of order 1 eV for the sum 
total of the masses. 
For the considered values of $\lsnd$ the 
sum of masses already exceeds the cosmological limit of $\simeq$ 1 eV. 
Therefore this scenario is highly constrained from cosmology in particular 
for higher values of $\lsnd$. 
The effective mass measured in neutrinoless 
double beta decay varies from 
$\sqrt{\Delta m^2_{\rm LSND}} \cos 2\theta_{\odot}$ to 
$\sqrt{\Delta m^2_{\rm LSND}}$. 
Since $\theta_{\odot} = \pi/4$ is no longer allowed by the current data, 
the minimum value of the effective mass is non-zero in this case\footnote{
This also is the case in scenario 2+2A.}. 
Depending on the value of $\Delta m^2_{\rm LSND}$, the lower value 
of \meff\ varies from $\simeq$ (0.04 -- 0.2) eV.
A part of this region is already disfavored by the limit 
from the Heidelberg-Moscow experiment.  Low values 
of $\sin^2 \theta_\odot$ and large values of \lsnd\ 
jeopardize scenarios 3+1Aa and 3+1Ab because 
the minimal \meff\ becomes too large.  
The neutrino mass measured in beta decay in this scenario is 
$\simeq \sqrt{\Delta m^2_{\rm LSND}}$ and coincides with the 
upper limit of the effective mass in neutrinoless double beta decay; 

\item the 
3+1B scenario corresponds to three neutrino states with a normal hierarchy 
separated from a fourth state by the LSND mass scale. 
For small $m_{1} < 0.01$ eV the sum of the masses is given by 
$\Sigma^{\rm 3+1B} \simeq \sqrt{\lsnd}$. Hence, the cases    
$\lsnd = 0.9$ eV$^2$ and \lsnd\ in the range (0.2 -- 0.5) eV$^2$ 
are consistent with the cosmological mass bound. 
Whether there is complete cancellation for the effective mass 
depends upon the product $|U_{e4}|^2 \sqrt{\lsnd}$. 
For our choice of parameters and hierarchical masses 
we got very low values of $\meff$ only if $\lsnd = 1.8$ eV$^2$. 
The mass measured in tritium beta decay is larger than $\meff$ for 
$m_1 \ls 0.1 $ eV;  

\item 
scenario 3+1C corresponds to the 
usual three generation inverted hierarchy picture plus an 
additional neutrino at a higher scale separated by the LSND gap. 
The second and the third state are quasi-degenerate at $\sqrt{\dma}$. 
For $m_1 > 0.1$ eV they become quasi-degenerate with the first state.  
The cosmological constraint on the 3+1C scenario is approximately the 
same as for 3+1B. 
The prediction for $\meff$ 
can be split into a contribution from a three neutrino inverted hierarchy 
scenario plus a term $e^{i \gamma} \, |U_{e4}|^2 \, \sqrt{\lsnd}$.  
The contribution of the three-flavor term to $\meff$ is 
between (0.007 -- 0.05) eV.
Hence, in this case we get very low values of $\meff$ for 
$\lsnd$ = 0.9 eV$^2$ and in the range (0.2 -- 0.5) eV$^2$. 
\end{itemize} 

Regarding the four neutrino mass and mixing matrices, 3+1 scenarios  
``embed'' in their mass and mixing matrices the well-known
three-flavor matrices. Corrections to these three-flavor matrices
are of order $\sqrt{\lsnd} \, \sin^2 \theta_{\rm LSND}$, i.e., of 
order $\dms$, for the mass matrix and of order
$\sin \theta_{\rm LSND}$ for the mixing matrix.
The 3+1B and 3+1C scenarios can in principle be motivated by an
approximate $L_e + L_\mu + L_\tau$ global symmetry\footnote{The 2+2 scenarios, 
discussed in the Appendix, have other candidates for global 
symmetries, for instance $L_e + L_\mu - L_\tau - L_s$. Moreover, 
there can be a $\tau$-$s$ exchange symmetry in analogy to the successful 
$\mu$-$\tau$ exchange symmetry for three neutrinos.}, whereas scenarios 
3+1Aa,b correspond to a $L_s$  global symmetry.  Scenario 3+1C 
can also be motivated by conservation of $L_e - L_\mu - L_\tau$.

\vspace{0.5cm}
\begin{center}
{\bf Acknowledgments}
\end{center}
The work of S.G.\ was supported by the Alexander-von-Humboldt-Foundation. 
The work of W.R.\ was supported by 
the ``Deutsche Forschungsgemeinschaft'' in the 
``Sonderforschungsbereich 375 f\"ur Astroteilchenphysik'' 
and under project number RO--2516/3--1.

\vskip 1cm {\it Note added}: 
The recent results from three year WMAP data in combination with 
large-scale structure and supernova data give a bound on the sum of three 
neutrino masses to be $\Sigma  \, m_\nu < 0.68$ eV (95\% C.L.) which is 
not very different from the first year limit \cite{wmap3}.
Therefore the bound on the 
sum of neutrino masses that we use in this paper is not expected to 
change significantly with this new data. 
More stringent bounds can be obtained by including the Lyman-$\alpha$ forest 
data \cite{seljak}. 

\begin{appendix}

\section{\label{appB}Oscillation Probabilities for Short Baseline 
Experiments} 
Here we list the relevant probabilities for short-baseline experiments in the
2+2 and 3+1 scenarios.  
The most general expression for neutrino survival or conversion 
probability  
for $N$ neutrino generations is given by 
\begin{equation}
P_{\nu_{\alpha}\nu_{\beta}} = \delta_{\alpha \beta} - 4~ \sum\limits_{j
> i}~ U_{\alpha i} ~U_{\beta i} ~U_{\alpha j} ~U_{\beta j}~
\sin^{2} \left(\frac{\pi L}{\lambda_{ij}} \right)
\label{pab}~,
\end{equation}
where $i, j$ varies from 1 to $N$ for $N$ generations, and 
$\lambda_{ij} = 2.47 \, ({E_{\nu}}/{\rm MeV})~
({{\rm eV}^{2}}/{\Delta m^2_{ij}}$) m. If the $\Delta m^2_{ij}$ 
corresponds to \dms, \dma\ or \lsnd, we denote it with 
$\lambda_{\odot}$, $\lambda_{\rm A}$ and $\lambda_{\rm LSND}$, respectively. 
The actual form of the various survival and transition
probabilities will depend on the spectrum of ${\Delta m}^{2}$ chosen. 
The above Eq.\ (\ref{pab}) is assuming the $CP$-phases to be zero. 
Below we list the relevant probabilities for short-baseline experiments in the
2+2 and 3+1 scenarios  for the reactor experiments Bugey 
\cite{bugey}, CHOOZ \cite{chooz} and the accelerator 
experiments CDHSW \cite{cdhsw}, LSND \cite{lsnd} and MiniBooNE 
\cite{miniboone}. 
The energy and length scales involved are such that one mass scale dominance 
approximation holds true in most of the following cases 
(excepting CHOOZ):  
\begin{itemize} 
\item Scenario 3+1Aa:  
\begin{equation}\nonumber
P_{\overline{\nu}_e\overline{\nu}_e} = 1 - 4~U_{e1}^2~(1 - U_{e1}^2)~ 
{\sin^2}({\pi L/\lambda_{\rm LSND}})~
~~~(\rm Bugey)
\label{b2}
\end{equation}
\begin{equation}\nonumber
P_{\overline{\nu}_e\overline{\nu}_e} = 1 - 2~U_{e1}^2 ~(1 - U_{e1}^2) - 4~
U_{e2}^2~ {\sin^2}({\pi L/\lambda_{\rm A}})
~~~(\rm CHOOZ)
\end{equation}
\begin{equation}\nonumber
P_{\overline{\nu}_{\mu}\overline{\nu}_{\mu}} = 1 - 4 ~U_{\mu 1}^2~(1
- U_{\mu1}^2)~ {\sin^2}({\pi L/\lambda_{\rm LSND}})~~~(\rm CDHSW)
\label{cdhsw2}
\end{equation}
\begin{equation}\nonumber
P_{\overline{\nu}_{\mu}\overline{\nu}_e} = 4 ~U_{e1}^2~ U_{\mu 1}^2 
~{\sin^2}({\pi L/\lambda_{\rm LSND}})~~~(\rm LSND, MiniBooNE)
\label{lsnd2}
\end{equation}
The relevant formulae for scenario 3+1Ab are obtained from the above 
by replacing $2 \leftrightarrow 4$. 
\item Scenario 3+1B:  
\begin{equation}\nonumber
P_{\overline{\nu}_e\overline{\nu}_e} = 1 - 4~U_{e4}^2~ (1 - U_{e4}^2)
~{\sin^2}({\pi L/\lambda_{\rm LSND}})
~~~(\rm Bugey)
\label{b12}
\end{equation}
\begin{equation}
P_{\overline{\nu}_e\overline{\nu}_e} = 1 - 2~U_{e4}^2~(1 - U_{e4}^2) -
4 ~U_{e3}^2
~{\sin^2}({\pi L/\lambda_{\rm A}})
~~~(\rm CHOOZ)
\label{b22}
\end{equation}
\begin{equation}\nonumber
P_{\overline{\nu}_{\mu}\overline{\nu}_{\mu}} = 1 - 4~ U_{\mu4}^2~(1
- U_{\mu4}^2)~{\sin^2}({\pi L/\lambda_{SBL}})~~~(\rm CDHSW)
\label{cdhsw22}
\end{equation}
\begin{equation}\nonumber
P_{\overline{\nu}_{\mu}\overline{\nu}_e} = 4~U_{e4}^2 ~U_{\mu 4}^2
~{\sin^2}({\pi L/\lambda_{\rm LSND}})~~~(\rm LSND, MiniBooNE)
\label{lsnd22}
\end{equation}
\end{itemize}
Note that the probabilities in the 3+1B picture can be obtained from 3+1Aa  
by replacing $U_{e1}$ by $U_{e4}$, 
$U_{\mu 1}$ by $U_{\mu 4}$ and $U_{e2}$
by $U_{e3}$ in the CHOOZ probability. 
The probabilities for the scenario 3+1C are the same as those 
of 3+1B excepting 
$U_{e3}^2$ in CHOOZ probability is to be replaced by $U_{e1}^2$. 
In the presence of $CP$ violation the mixing matrix elements are 
complex. In the above 3+1 oscillation probabilities one then has to 
replace $U_{\alpha i}^2$ with $|U_{\alpha i}|^2$.\\ 

Let us for the sake of completeness also give the oscillation 
probabilities in the 2+2 schemes: 
\begin{itemize} 
\item Scenario 2+2A: 
\begin{equation} \nonumber
P_{\overline{\nu}_e\overline{\nu}_e} = 1 - 4 ~(U_{e1}^2 + U_{e2}^2)~
 (1 - U_{e1}^2 - U_{e2}^2)~ {\sin^2}({\pi L/\lambda_{\rm LSND}}) 
~~~(\rm Bugey)
\label{b22a}
\end{equation}
\begin{equation}\nonumber
P_{\overline{\nu}_e\overline{\nu}_e} = 1 - 2 ~(U_{e1}^2 + U_{e2}^2) ~
(1 - U_{e1}^2 - U_{e2}^2) - 4 ~U_{e1}^2 ~U_{e2}^2~ 
{\sin^2}({\pi L/\lambda_{\rm A}})~~~(\rm CHOOZ)
\label{c22a}
\end{equation}
\begin{equation}\nonumber
P_{\overline{\nu}_e\overline{\nu}_e} = 1 - 4 ~(U_{\mu 1}^2 + U_{\mu 2}^2)~
 (1 - U_{\mu 1}^2 - U_{\mu 2}^2)~ {\sin^2}({\pi L/\lambda_{\rm LSND}})  
~~~(\rm CDHSW)
\label{cd22a}
\end{equation}
\begin{equation}\nonumber
P_{\overline{\nu}_{\mu}\overline{\nu}_e} = 
4~(U_{e1} ~U_{\mu 1}+ U_{e 2} ~U_{\mu 2 })^2~ 
{\sin^2}({\pi L/\lambda_{\rm LSND}})~~~(\rm LSND, MiniBooNE)
\label{lsnd22a}
\end{equation}
\end{itemize}
The survival and oscillation probabilities in scenario 
2+2B are obtained from those of scenario 2+2A 
by making the change $1 \leftrightarrow 3$ and $2 \leftrightarrow 4$. 
In case there are Dirac phases in the PMNS matrix, again 
$U_{\alpha i}^2$ will have to be replaced with $|U_{\alpha i}|^2$. 
In what regards the (LSND, MiniBooNE) probability, the relevant term would 
read $|U_{e1}^\ast ~U_{\mu 1} + U_{e 2}^\ast ~U_{\mu 2 }|$ instead of 
$(U_{e1} ~U_{\mu 1}+ U_{e 2} ~U_{\mu 2 })$.

\newpage

\begin{figure}[ht]
\begin{center}
\vspace{-3cm}
\epsfig{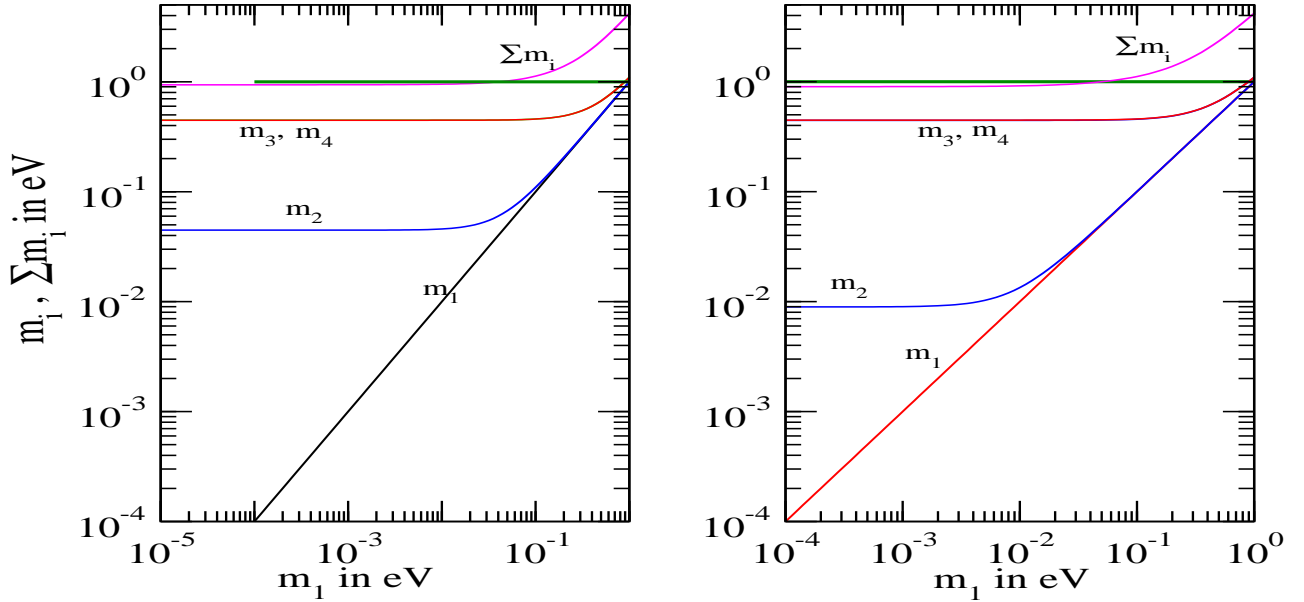}
\caption{\label{fig:2+2_mass}All four neutrino masses and their
sum $\Sigma$ against the smallest neutrino mass $m_1$ for scenario
2+2A (left) and 2+2B (right). The masses $m_3$ and $m_4$ are not
distinguishable in the plot. We choose \lsnd = 0.2 eV$^2$.}
\end{center}

\section{\label{sec:appa}Neutrino Masses and 
Neutrinoless Double Beta Decay in 2+2 Scenarios}
The 2+2 scenarios are disfavored by the combination of 
solar and atmospheric data regardless if LSND results are 
confirmed by MiniBooNE or not. 
However, for the sake of completeness we discuss in this Section 
the constraints on 2+2 mass spectra from cosmology, neutrinoless 
double beta decay and tritium beta decay. 
\subsection{\label{sec:0vbb2+2a}Neutrino Masses and
Neutrinoless Double Beta Decay in Scenario 2+2A}

\end{figure}
\begin{figure}[hbt]
\centerline{\psfig{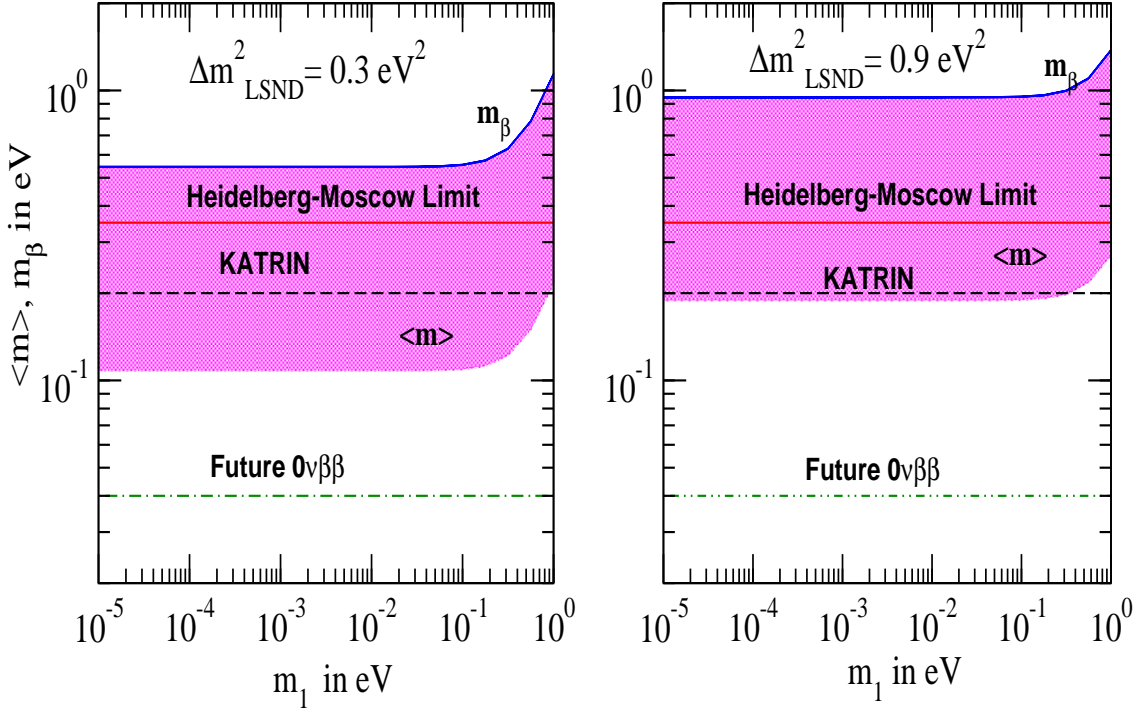}}
\caption{The effective mass  
$\meff$ and the mass $m_{\beta}$ measured in beta decay experiments 
plotted 
as a function of the lowest mass in 2+2A
scenario. The left hand column is for $\Delta m^2_{\rm LSND}$ = 0.3 eV$^2$
and $U_{e1}^2 = U_{e2}^2 = 0.002$. 
The right hand column is for $\Delta m^2_{\rm LSND}$ = 0.9 eV$^2$
and $U_{e1}^2 = U_{e2}^2 = 0.005$. 
The other parameters are varied in their current 3$\sigma$ allowed range
and all the phases are varied between 0 and $2\pi$. 
Also shown is the mass $m_{\beta}$ that will be 
measured in beta decay experiments, the current and a prospective future 
limit on the effective mass and the future KATRIN limit.
}
\label{fig:2+2a0vbb}
\end{figure}

On the left side of Fig.\ \ref{fig:2+2} the mass ordering of scheme
2+2A can be seen. In this scheme it holds that
$\dms = \Delta m^2_{43}$ and $\dma = \Delta m^2_{21}$.
One has 
\bea
m_2 = \sqrt{m_1^2 + \dma}   ~,\\[0.2cm]
m_3 = \sqrt{m_1^2 + \Delta m^2_{\rm LSND} - \dms}~,\\[0.2cm]
m_4 = \sqrt{m_1^2 + \lsnd }~.
\label{mass:2+2a}
\eea
The left side of Fig.\ \ref{fig:2+2_mass} shows the
four mass values and their sum $\Sigma$
as a function of the smallest mass $m_1$. Due to $\lsnd \gg \dma$ the
masses $m_4$ and $m_3$ are hardly
distinguishable in the plot and for $m_1 \gs 0.1$ eV the two lightest
masses $m_1$ and $m_2$ become quasi-degenerate.
To clarify the role of the cosmology bounds on $\Sigma$,
we included in the Figure a value of $\Sigma = 1$ eV. To let
the sum of neutrino masses lie below this limit, small values
of \lsnd{} and of $m_1$ are implied:
for small $m_1 \ls 0.05$ eV it holds that
\be
m_4 \simeq m_3 \simeq \sqrt{\lsnd} \gg m_{2,1}~.
\ee
Hence, $\Sigma \simeq 2\sqrt{\lsnd}$ and therefore low values of
$\lsnd \ls 0.3$ eV$^2$ are implied by the condition $\Sigma \ls 1$ eV.

Turning to the constraints on the mixing matrix elements,
$|U_{e1}|^2$ and $|U_{e2}|^2$ are constrained by the
short baseline reactor experiment Bugey and the reactor 
experiment CHOOZ. 
$\sin^2 2 \theta_{\rm LSND}$ is given by
the combination $ (U_{e1} U_{\mu1} + U_{e2} U_{\mu 2})^2$ and the combination
$(|U_{\mu 1}|^2 + |U_{\mu 2}|^2) $ will be constrained by the CDHS
experiment.
We give the relevant expressions for the probability in 
Appendix \ref{appB}. 
Since the one mass scale dominance approximation holds, one can use 
two-parameter plots to find the constraints on these mixing 
parameters. However, for the sake of illustration in this paper we use 
the values of $\Delta m^2_{\rm LSND}$ and $\sin^22\theta_{\rm LSND}$ 
from the MiniBooNE sensitivity plot given, e.g., in \cite{sorel}. 
We take the representative values\footnote{It is to be noted that 
the constraints on mixing angles 
from SBL experiments are not as severe in the 2+2 case as in the 
3+1 case.} for  
($\lsnd$,~$\sin^22\theta_{\rm LSND}$) as (0.3,0.02) and (0.9,0.008).
To extract $U_{e1}$ and $U_{e2}$ from $\sin^2 2\theta_{\rm LSND}$ we make 
the plausible assumption $U_{e1} \simeq U_{e2}$ and 
$U_{\mu1 }^2 = U_{\mu2}^2 =0.5$ as implied by atmospheric data. 
This assumption was for instance used in \cite{4vbb}. 
With this assumption we have $U_{e1}^2 = U_{e2}^2 = 0.002~(0.005)$ for 
\lsnd = 0.9 (0.3) eV$^2$.
In any case it is to be noted that in the 2+2A scenario $U_{e1}$ and $U_{e2}$ 
multiply the smaller masses $m_1$ and $m_2$, and as we will see below 
their contribution to effective mass as well as the mass measured in 
beta decay is sub-leading. 
We furthermore have
$|U_{e3}| \simeq \cos \theta_\odot$ and 
$|U_{e4}| \simeq \sin \theta_\odot$.

Since $m_1$ and $m_2$ are small in scenario 2+2A and in addition
multiplied with the small elements $|U_{e1}|^2$ and $|U_{e2}|^2$,
respectively, we can neglect terms including these quantities in what
follows.
Then the effective mass in scenario 2+2A reads
\be
\meff^{\rm 2+2A} \simeq \sqrt{\lsnd} \, \sqrt{1 - \sin^2 2\theta_\odot \,
\sin^2 (\beta - \gamma)/2}~.
\ee
The non-maximality of solar neutrino mixing implies therefore a
non-vanishing effective mass. This is in analogy to the three-flavor
case with an inverted hierarchy or quasi-degenerate neutrinos. 
Choosing for instance $\sqrt{\lsnd} \simeq 0.5$ eV and the values of
$\theta_\odot$ from Eq.\ (\ref{eq:sssrange}), we can predict that 
\be
\cos 2 \theta_\odot \, \sqrt{\lsnd} \simeq 0.1 \eV
\ls \meff^{\rm 2+2A} \ls \sqrt{\lsnd} \ls 0.5 \eV~.
\ee
This range of \meff{} is well within reach of currently running or planned
\obb{} experiments. In Fig.\ \ref{fig:2+2a0vbb} we show the effective mass
as a function of the smallest neutrino mass $m_1$.
We also show the present bound from the Heidelberg-Moscow experiment
in this Figure, together with a prospective future limit. 
It is to be noted that some part of the  
regions are already disfavored by the Heidelberg-Moscow limit. 
Thus non-maximality of solar neutrino mixing angle  coupled with 
the existing limit from Heidelberg-Moscow experiment 
already puts some constraint on the 2+2A mass pattern.

Neglecting terms proportional to $U_{e1}^2 \, m_1$ and $U_{e2}^2 \, m_2$,
we have for the kinematic neutrino mass
\be
m_{\beta}^{\rm 2+2A} \simeq \sqrt{ |U_{e3}|^2 \, m_3^2 + |U_{e4}|^2 \, m_4^2 }
\simeq \sqrt{\lsnd}~.
\ee
Since cosmology implies that $\sqrt{\lsnd}$ is below roughly
0.3 eV, we expect that $m_{\beta}$ should be close to the lowest value
reachable by KATRIN, but close to the current limit on the sum of neutrino
masses from cosmology.
Since
\be
m_{\beta}^{\rm 2+2A} \simeq \sqrt{\lsnd} \simeq \Sigma^{\rm 2+2A}/2
\simeq \sqrt{\lsnd}  \, \sqrt{1 - \sin^2 2\theta_\odot \,
\sin^2 (\beta - \gamma)/2}~,
\ee
one can in
principle obtain a set of consistency checks of scenario 2+2A,
and obtain some information on the Majorana phase combination
$\beta - \gamma$.

\subsection{\label{sec:0vbb2+2b}Neutrino Masses and
Neutrinoless Double Beta Decay in Scenario 2+2B}

The right panel of Fig.\ \ref{fig:2+2_mass} shows the
values of the four neutrino masses and their sum $\Sigma$ in scenario 2+2B,
in which one has $\dms = m_2^2 - m_1^2$ and $\dma = m_4^2 - m_3^2$.
We can write the masses of the neutrino states in terms of the smallest 
mass $m_1$ and the three mass squared differences as 
\bea
~m_2 = \sqrt{m_1^2 + \dms}   ~,\\[0.2cm]
m_3 = \sqrt{m_1^2 + \Delta m^2_{\rm LSND} - \Delta m^2_{\rm A}}~,\\[0.2cm]
m_4 = \sqrt{m_1^2 + \lsnd ~}~.
\label{mass:2+2b}
\eea
For the mass values similar statements as for scenario 2+2A apply, namely
that $m_4 \simeq m_3 \simeq \sqrt{\lsnd} \simeq \Sigma^{\rm 2+2B}/2$,
where \lsnd{} should lie at the
lower end of its allowed range in order to obey the constraints
from cosmology. The smallest neutrino mass $m_1$ should lie below 
0.05 eV and therefore $m_2$ and $m_1$ are typically not very close 
to each other,
unless $m_1 \simeq 0.02 - 0.05$ eV.

In what regards the mixing matrix elements, scenario 2+2B is obtained
from scenario 2+2A by exchanging the indices
$1 \leftrightarrow 3$ and $2 \leftrightarrow 4$.
Hence, $|U_{e1}|$ and $|U_{e2}|$ are roughly given by $\cos \theta_\odot$
and $\sin \theta_\odot$, respectively.
The elements $|U_{e3}|$ and $|U_{e4}|$, however,
are implied to be small.

\begin{figure}\vspace{-2cm}
\centerline{\psfig{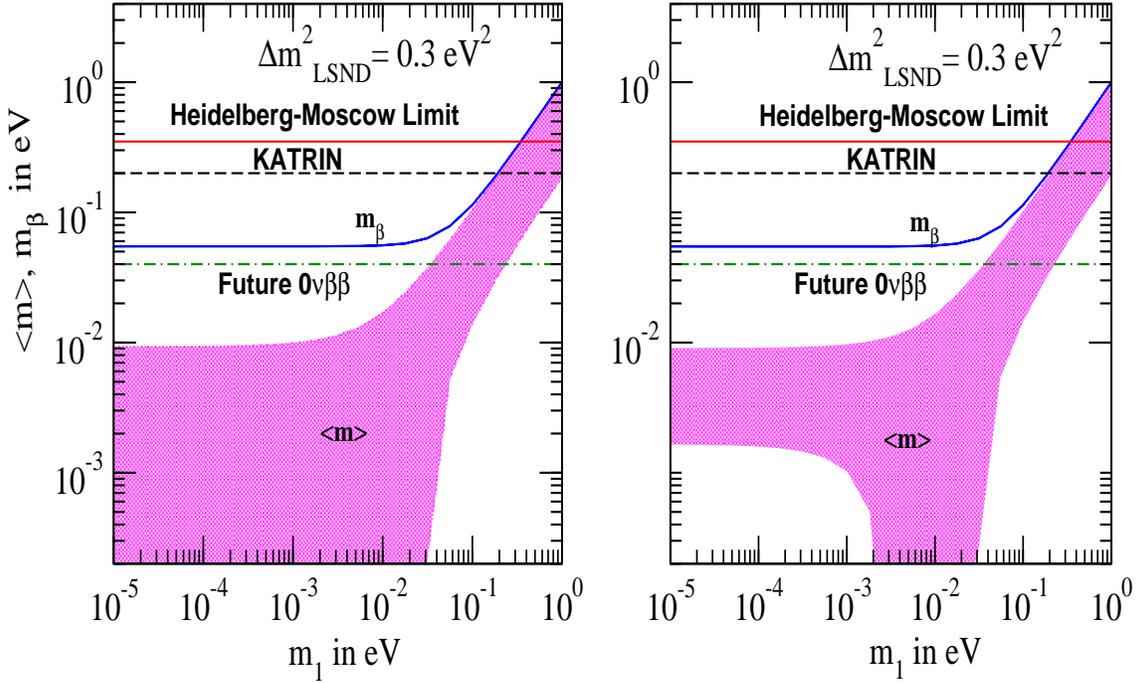}}
\caption{Same as previous Figure for scenario 2+2B 
and $\Delta m^2_{\rm LSND}$ = 0.3 eV$^2$. 
The left hand column is for $U_{e3}^2 = U_{e4}^2 = 0.005$, 
the right hand column is for $U_{e4}^2 = 0 $ and $U_{e3}^2 = 0.01$. 
}
\label{fig:2+2b0vbb}
\end{figure}

We show in Fig.\ \ref{fig:2+2b0vbb} the effective mass as a function of the
smallest mass.
The effective mass is approximately  given by
\bea
\meff^{\rm 2+2B} \simeq \left|
\cos^2 \theta_\odot \, m_1 + \sin^2 \theta_\odot \, m_2 \, e^{i \alpha}
+ |U_{e3}|^2 \, m_3 \, e^{i \beta} + |U_{e4}|^2 \, m_4 \, e^{i \gamma}
\right| \\[0.2cm]
\simeq
\left|
\cos^2 \theta_\odot \, m_1 + \sin^2 \theta_\odot \, m_2 \, e^{i \alpha}
+ \sqrt{\lsnd} \left( |U_{e3}|^2 \, e^{i \beta} + |U_{e4}|^2 \, e^{i \gamma}
\right)
\right| \\[0.2cm]
\simeq \left| \sin^2 \theta_\odot \, \sqrt{\dms} +
\sqrt{\lsnd} \left( |U_{e3}|^2 \, \, e^{i (\beta - \alpha)}
+ |U_{e4}|^2 \, e^{i (\gamma - \beta)}
\right) \right|~,
\eea
where we neglected $m_1$ for the last approximation.

Since $U_{e3}^2$ and $U_{e4}^2$ are small,
the two large masses $m_3$ and $m_4$ are
multiplied with small mixing matrix elements, whereas the
small masses $m_1$ and $m_2$ are multiplied with large mixing matrix
elements. As a consequence, there can be 
cancellations leading to
a very small or zero effective mass.
Note the analogy of this situation with the three-flavor case:
in the inverted hierarchy the large masses are multiplied
with mixing matrix elements corresponding to the large solar neutrino mixing,
whose non-maximality allows no cancellation. In the normal hierarchy,
the largest mass $m_3$ is multiplied with the smallest mixing
matrix element, and
complete cancellation can occur.
Since the degree of cancellation depends on the values of the two 
small quantities 
$U_{e3}^2$ and $U_{e4}^2$ 
for this case, we present our 
results for $\lsnd$~= 0.3 eV$^2$ and two different choices for 
$U_{e3}^2$ and $U_{e4}^2$. In the left panel we show the plot 
where we assume $U_{e3}^2 \simeq U_{e4}^2 = 0.005$.
In this case complete cancellation 
can occur. In the right panel we present our results with 
$U_{e4}^2 = 0$ and $U_{e3}^2 = 0.01$. 
In this case the minimal $\meff$ has a higher value. 
The two terms
in the last equation have the typical order
$\sqrt{\dms} \sin^2 \theta_\odot \simeq (2-4)\cdot 10^{-3}$ eV
and $\sqrt{\lsnd} |U_{e3}|^2  \simeq 0.005$ eV. 
The upper limit on $\meff^{\rm 2+2B}$ is then roughly given by 0.01 eV. 
Hence, we can in principle distinguish scenario 2+2A from 2+2B via \obb{}
as long as $m_1$ is small. This
is analogous to the situation normal vs.\ inverted hierarchy in
the three flavor case.

The kinematic neutrino mass as measurable in the KATRIN experiment
is given by
\bea
m_{\beta}^{\rm 2+2B} =
\sqrt{|U_{e1}|^2 \, m_1^2 + |U_{e2}|^2 \, m_2^2 + |U_{e3}|^2 \, m_3^2 +
|U_{e4}|^2 \, m_4^2 } \\[0.2cm]
\simeq \sqrt{\cos^2 \theta_\odot \, m_1^2 + \sin^2 \theta_\odot \, m_2^2
+ \lsnd \, \left( |U_{e3}|^2 + |U_{e4}|^2 \right)} \\[0.2cm]
\simeq \sqrt{\dms \, \sin^2 \theta_\odot + \lsnd \,
\left( |U_{e3}|^2 + |U_{e4}|^2 \right)}~,
\eea
where we neglected again $m_1$ in the last approximation.
Both terms are of similar magnitude and we can expect that
$m_{\beta}^{\rm 2+2B} \simeq 0.1$ eV, larger than the effective mass by an
order of magnitude and below the future KATRIN limit.

\section{\label{sec:mnu2+2}The Mass Matrix in the 2+2 Scenarios}
Now we discuss the form of the mass matrices in 2+2 scenarios. Our 
approach and the approximations made are the same as for the 3+1 case, 
and details are given in Section \ref{sec:matrices}.

\subsection{\label{sec:mnu2+2a}The Mass Matrix in Scenario 2+2A}
For scenario 2+2A we can express the mixing matrix as follows 
\cite{4mix1,4mix2}:
\be \label{eq:U2+2A}
U^{\rm 2+2A} \simeq
\left(
\bav
\lambda & \lambda & \cos \theta_\odot & \sin \theta_\odot \\
\cos \theta_{\rm atm} & \sin  \theta_{\rm atm} & \lambda & \lambda  \\
-\sin \eta\, \sin  \theta_{\rm atm} & \sin \eta\, \cos \theta_{\rm atm} &
-\cos \eta\, \sin  \theta_\odot & \cos \eta\, \cos  \theta_\odot \\
-\cos \eta\, \sin  \theta_{\rm atm} &  \cos \eta\, \cos \theta_{\rm atm} &
\sin \eta\, \sin  \theta_\odot & -\sin \eta\, \cos \theta_\odot
\ea
\right)\, P~.
\ee
The parameter $\eta$ indicates inasmuch sterile neutrinos
participate in atmospheric or solar neutrino oscillations.
For $\eta=0$ atmospheric neutrinos
oscillate completely into sterile ones and for $\eta=\pi/2$ solar
neutrinos oscillate into sterile ones.
With a given mass hierarchy we can obtain now the approximate form of the
mass matrix. 
Glancing at Fig.\ \ref{fig:2+2_mass}, we identify two
interesting possibilities, namely
\be
\baz
\mbox{(i)} &  \sqrt{\lsnd} \simeq
m_4 \simeq m_3 \gg m_2 \simeq \sqrt{\dms} \gg m_1 ~,\\[0.2cm]
\mbox{(ii)} & \sqrt{\lsnd} \simeq
m_4 \simeq m_3 \gg m_2 \simeq m_1 \simeq 0.1 \eV~.
\ea
\ee
The first case (i)
corresponds to a very small mass $m_1$ and the second one (ii) to two
quasi-degenerate pairs, though only a small range of $m_1$ values
allows for this possibility. Since the form of
$m_\nu$ is similar in both cases, we mainly discuss
case (i).
We have then $m_2 \simeq m_4 \, \lambda^2$ and the mass matrix reads
\be
m_\nu^{\rm 2+2A} \sim \sqrt{\lsnd} \,
\left(
\bav
e^{i \beta} \, c_\odot^2 + e^{i \gamma} \, s_\odot^2 & \lambda &
c_\eta \left(e^{i \gamma} -  e^{i \beta} \right) &
s_\eta \left(e^{i \gamma} -  e^{i \beta} \right) \\[0.2cm]
\cdot & 0 & c_\eta \, \lambda & s_\eta \, \lambda \\[0.2cm]
\cdot & \cdot & c_\eta^2
\left(e^{i \gamma} \, c_\odot^2 +  e^{i \beta} \, s_\odot^2\right) &
c_\eta s_\eta
\left(e^{i \gamma} \, c_\odot^2 +  e^{i \beta} \, s_\odot^2\right) \\[0.2cm]
\cdot & \cdot & \cdot & s_\eta^2
\left(e^{i \gamma} \, c_\odot^2 +  e^{i \beta} \, s_\odot^2\right)
\ea
\right).
\ee
We defined the obvious notation $c_\odot = \cos \theta_\odot$ and
$s_\odot = \sin \theta_\odot$. Terms of order $\lambda^2$ and unimportant
factors, such as a coefficient $c_1 \, e^{i \gamma} \, c_\odot -
c_2 \, e^{i \beta} \, s_\odot$ for the $\mu \tau$ element,
are not included in our expressions. The factors $c_{1,2}$ depend on the
precise values of the CHOOZ and the LSND angles.
Note that contributions of the atmospheric mixing are suppressed
in the mass matrix.
If the heavy states $m_3$ and $m_4$ have equal $CP$ parities, or when
$\beta = \gamma$, then
this leads to the vanishing of the $e\tau$ and $e s$ entries of $m_\nu$,
independent of $\eta$.
In case of opposite $CP$ parities of $\nu_3$ and $\nu_4$ (which would imply
enhanced stability with respect to radiative corrections), the
$ee$ element and the $\tau s$ block of $m_\nu$ would be slightly
suppressed by a factor $\cos 2 \theta_\odot$.
Assuming as yet another approximation that $\theta_\odot = \pi/4$ 
would make these entries vanish.
Several special cases can be obtained from the above matrix.
For instance, if solar neutrinos oscillate entirely in sterile neutrinos,
i.e., $\eta = \pi/2$, then
\be
m_\nu^{\rm 2+2A} \sim \sqrt{\lsnd} \,
\left(
\bav
e^{i \beta} \, c_\odot^2 + e^{i \gamma} \, s_\odot^2  & 0 & 0 &
e^{i \gamma}  - e^{i \beta} \\[0.2cm]
\cdot & 0 & 0 & 0 \\[0.2cm]
\cdot & \cdot & 0 & 0 \\[0.2cm]
\cdot & \cdot & \cdot &
e^{i \gamma} \, c_\odot^2 + e^{i \beta} \, s_\odot^2
\ea
\right)~.
\ee
This matrix conserves the flavor charge $L_\mu + L_\tau$.
If we choose equal $CP$ parities of $\nu_3$ and $\nu_4$, or when
$\beta = \gamma$, then the $e\tau$ entry vanishes.

In analogy, if atmospheric neutrinos oscillate entirely in
sterile neutrinos ($\eta = 0$) then one finds
\be
m_\nu^{\rm 2+2A} \sim \sqrt{\lsnd} \,
\left(
\bav
e^{i \beta} \, c_\odot^2 + e^{i \gamma} \, s_\odot^2  & 0 &
e^{i \gamma}  - e^{i \beta} & 0 \\[0.2cm]
\cdot & 0 & 0 & 0 \\[0.2cm]
\cdot & \cdot &   e^{i \gamma} \, c_\odot^2 + e^{i \beta} \, s_\odot^2
& 0 \\[0.2cm]
\cdot & \cdot & \cdot & 0
\ea
\right)~.
\ee
Opposite $CP$ parities (a Pseudo-Dirac structure) of
$\nu_3$ and $\nu_4$ will again lead to $(m_\nu)_{e \tau} = 0$.

Now consider $c_\eta \simeq s_\eta$. The mass matrix takes the form
\be
m_\nu^{\rm 2+2A} \sim \sqrt{\lsnd} \,
\left(
\bav
e^{i \beta} \, c_\odot^2 + e^{i \gamma} \, s_\odot^2 & \lambda &
\left(e^{i \gamma} -  e^{i \beta} \right) &
\left(e^{i \gamma} -  e^{i \beta} \right) \\[0.2cm]
\cdot & 0 & \lambda & \lambda \\[0.2cm]
\cdot & \cdot &
\left(e^{i \gamma} \, c_\odot^2 +  e^{i \beta} \, s_\odot^2\right) &
\left(e^{i \gamma} \, c_\odot^2 +  e^{i \beta} \, s_\odot^2\right) \\[0.2cm]
\cdot & \cdot & \cdot &
\left(e^{i \gamma} \, c_\odot^2 +  e^{i \beta} \, s_\odot^2\right)
\ea
\right)~.
\ee
We therefore find an approximate $\tau$-$s$ symmetry, in analogy to the
successful $\mu$-$\tau$ symmetry of the three neutrino case \cite{mutau}.
It is present when sterile neutrinos participate equally in solar
and atmospheric neutrino oscillations. The $\tau$-$s$ symmetry does
strictly speaking only say that $(m_\nu)_{\tau \tau} = (m_\nu)_{ss}$, here it
holds in addition that $(m_\nu)_{\tau s} = (m_\nu)_{ss}$.
If we consider the matrix
\be
m_\nu =
\left(
\bav
a & b & d & d \\[0.2cm]
\cdot & h & e & e \\[0.2cm]
\cdot & \cdot & f & f \\[0.2cm]
\cdot & \cdot & \cdot & f
\ea
\right)~,
\ee

we see that one eigenvalue is zero. Setting for simplicity
$b = h = e = 0$ (these entries are suppressed in the previous
equation), leads to two vanishing mass eigenvalues and
$m_{3,4} = \frac{1}{2} (a + 2 f \mp \sqrt{8 d^2 + (a - 2 f)^2})$,
and therefore $\dms = (a + 2 f) \, \sqrt{8 d^2 + (a - 2 f)^2}$.
In this limit the atmospheric $\Delta m^2$ is vanishing.
We have $U_{e1} = U_{e2}=0$, $|U_{e3}|^2 =
\frac{1}{2} - (a/2 - f)/(\sqrt{8 d^2 + (a - 2 f)^2})$ and
$|U_{\mu 1}| = |U_{\tau 1}| = 1/\sqrt{2}$. Hence, the approximate
form of Eq.\ (\ref{eq:U2+2A}) is almost reproduced when
$a \simeq -2 f$. Small breaking
terms can in principle help to reach full agreement.

\noindent
The second interesting case (ii) occurs when 
$m_4 \simeq m_3 \gg m_2 \simeq m_1 \simeq \lambda \, m_4 \simeq 0.1$ eV.
The implications are similar to case (i), but for completeness we give
the resulting form of the mass matrix:
\be
m_\nu^{\rm 2+2A} \sim \sqrt{\lsnd} \,
\left(
\bav
e^{i \beta} \, c_\odot^2 + e^{i \gamma} \, s_\odot^2 & \lambda &
c_\eta \left(e^{i \gamma} -  e^{i \beta} \right) &
s_\eta \left(e^{i \gamma} -  e^{i \beta} \right) \\[0.2cm]
\cdot & \lambda & s_\eta \, c_{\rm atm} \, s_{\rm atm} & s_\eta   \\[0.2cm]
\cdot & \cdot & c_\eta^2
\left(e^{i \gamma} \, c_\odot^2 +  e^{i \beta} \, s_\odot^2\right) &
c_\eta s_\eta
\left(e^{i \gamma} \, c_\odot^2 +  e^{i \beta} \, s_\odot^2\right) \\[0.2cm]
\cdot & \cdot & \cdot & s_\eta^2
\left(e^{i \gamma} \, c_\odot^2 +  e^{i \beta} \, s_\odot^2\right)
\ea
\right).
\ee
Comparing with case (i), we see that the second row of the mass
matrix differs. It vanishes to first order when $\eta=0$.
If $\eta \neq 0$ the
atmospheric neutrino mixing angle has some dependence on the form of
the mass matrix.
As an additional approximation, let us take $\theta_\odot = \pi/4$.
Then, for $e^{i \gamma} +  e^{i \beta} = 0$, i.e., a Pseudo-Dirac
structure of the two heavy masses, and $c_\eta \simeq s_\eta$, we have
\be
m_\nu^{\rm 2+2A} \sim \sqrt{\lsnd} \,
\left(
\bav
0 & 0 & 1 & 1 \\[0.2cm]
\cdot & 0 & 1 & 1 \\[0.2cm]
\cdot & \cdot & 0 & 0 \\[0.2cm]
\cdot & \cdot & \cdot & 0
\ea
\right)~.
\ee
This matrix conserves $L_e + L_\mu - L_\tau - L_s$.
Indeed, this global symmetry has been used in \cite{sruba} and, in somewhat
different form in \cite{rabi_babu}, to explain the neutrino
data including LSND.
Moreover, the above matrix has all diagonal entries zero, a property
typically shared by radiative models of neutrino mass generation.
In Ref.\ \cite{roy_sudhir} such a case is treated.

\subsection{\label{sec:mnu2+2b}The Mass Matrix in Scenario 2+2B}
For scenario 2+2B we can express the mixing matrix by exchanging in the
mixing matrix from scheme 2+2A the indices $1 \leftrightarrow 3$ and
$2 \leftrightarrow 4$.
Hence,
\be
U^{\rm 2+2B} \simeq
\left(
\bav
\cos \theta_\odot & \sin \theta_\odot & \lambda & \lambda \\
\lambda & \lambda & \cos \theta_{\rm atm} & \sin  \theta_{\rm atm} \\
-\cos \eta\, \sin  \theta_\odot & \cos \eta\, \cos  \theta_\odot &
-\sin \eta\, \sin  \theta_{\rm atm} & \sin \eta\, \cos \theta_{\rm atm}\\
\sin \eta\, \sin  \theta_\odot & -\sin \eta\, \cos \theta_\odot &
-\cos \eta\, \sin  \theta_{\rm atm} &  \cos \eta\, \cos \theta_{\rm atm}
\ea
\right)\, P~.
\ee
Here we again put terms of order $\lambda \sim 0.1$, which is the
typical order of both the CHOOZ angle and the LSND parameter.
We have again two cases of interest:
\be
\baz
\mbox{(i)} &  \sqrt{\lsnd} \simeq
m_4 \simeq m_3 \gg m_2 \simeq \sqrt{\dms} \gg m_1 ~,\\[0.2cm]
\mbox{(ii)} & \sqrt{\lsnd} \simeq
m_4 \simeq m_3 \gg m_2 \simeq m_1 \simeq 0.1 \eV~.
\ea
\ee
Let us start with case (i), for which
\be \nonumber
m_\nu^{\rm 2+2B} \sim \sqrt{\lsnd} \,
\left(
\bav
0 &  \lambda & s_\eta \, \lambda  & c_\eta \, \lambda \\[0.2cm]
\cdot &  c_{\rm atm}^2 \, e^{i \beta} + s_{\rm atm}^2 \, e^{i \gamma} &
c_{\rm atm} s_{\rm atm} s_\eta (e^{i \beta} - e^{i \gamma}) &
c_{\rm atm}  s_{\rm atm}  c_\eta (e^{i \beta} - e^{i \gamma})
\\[0.2cm]
\cdot & \cdot & s_\eta^2 (c_{\rm atm}^2 \, e^{i \gamma} + s_{\rm atm}^2 \, e^{i \beta}) &
c_\eta s_\eta  (c_{\rm atm}^2 \, e^{i \gamma} + s_{\rm atm}^2 \, e^{i \beta}) \\[0.2cm]
\cdot & \cdot & \cdot &
c_\eta^2 (c_{\rm atm}^2 \, e^{i \gamma} + s_{\rm atm}^2 \, e^{i \beta})
\ea
\right)
\ee
In contrast to scenario 2+2A it is the solar neutrino mixing angle
whose contribution to the mass matrix is suppressed.
Note again the approximate $\tau$-$s$ exchange symmetry of the
mass matrix in case of $s_\eta \simeq c_\eta$.
The (close-to-)maximality of $\theta_{\rm atm}$ allows to further
simplify the mass matrix to
\be
m_\nu^{\rm 2+2B} \sim \sqrt{\lsnd} \,
\left(
\bav
0 &  \lambda & s_\eta \, \lambda  & c_\eta \, \lambda \\[0.2cm]
\cdot &  e^{i \beta} +  e^{i \gamma} &
s_\eta (e^{i \beta} - e^{i \gamma}) &
 c_\eta (e^{i \beta} - e^{i \gamma})
\\[0.2cm]
\cdot & \cdot & s_\eta^2 (e^{i \gamma} +e^{i \beta}) &
c_\eta s_\eta  (e^{i \gamma} + e^{i \beta}) \\[0.2cm]
\cdot & \cdot & \cdot &
c_\eta^2 (e^{i \gamma} + e^{i \beta})
\ea
\right)~.
\ee
Opposite (identical) $CP$ parities of $\nu_3$ and $\nu_4$ lead
to a vanishing $\mu\mu$ entry and $\tau s$ block
($\mu \tau$ and $\mu s$ elements). If we indeed impose a Pseudo-Dirac
structure on $\nu_3$ and $\nu_4$, then we have
\be
m_\nu^{\rm 2+2B} \sim \sqrt{\lsnd} \,
\left(
\bav
0 & \lambda & s_\eta \, \lambda  & c_\eta \, \lambda \\[0.2cm]
\cdot & 0 & s_\eta & c_\eta \\[0.2cm]
\cdot & \cdot & 0 & 0 \\[0.2cm]
\cdot & \cdot & \cdot & 0
\ea
\right)~.
\ee
If $c_\eta \simeq s_\eta$ and the $e\mu$ entry of $m_\nu$ (recall that
there can be a coefficient) is more suppressed than the $e\tau$ and $e s$
elements, then we have again
a mass matrix conserving $L_e + L_\mu - L_\tau - L_s$.
\noindent
Setting $\eta = 0$ (atmospheric-sterile oscillations) and
$\theta_{\rm atm}=\pi/4$, then at leading order
\be
m_\nu^{\rm 2+2B} \sim \sqrt{\lsnd} \,
\left(
\bav
0 & 0 & 0  & 0 \\[0.2cm]
\cdot &  e^{i \beta} +  e^{i \gamma} & 0
& e^{i \beta} - e^{i \gamma} \\[0.2cm]
\cdot & \cdot & 0 & 0 \\[0.2cm]
\cdot & \cdot & \cdot & e^{i \beta} +  e^{i \gamma}
\ea
\right)~,
\ee
conserving $L_e + L_\tau$. Recall that $\eta=\pi/2$ in scenario 2+2A lead to
conservation of $L_\mu + L_\tau$.
By choosing opposite or identical $CP$ parities one can further
simplify the mass matrix.
\noindent
If $\eta = \pi/2$ (solar-sterile oscillations), then
at leading order
\be
m_\nu^{\rm 2+2B} \sim \sqrt{\lsnd} \,
\left(
\bav
0 & 0 & 0  & 0 \\[0.2cm]
\cdot &  e^{i \beta} +  e^{i \gamma} & e^{i \beta} - e^{i \gamma}
& 0 \\[0.2cm]
\cdot & \cdot & e^{i \beta} +  e^{i \gamma} & 0 \\[0.2cm]
\cdot & \cdot & \cdot & 0
\ea
\right)~,
\ee
conserving $L_e + L_s$.\\
\noindent
Finally, we note that case (ii), defined by
$m_4 \simeq m_3 \gg m_2 \simeq m_1 \simeq \lambda \, m_4 \simeq 0.1$ eV,
leads to
\bea
m_\nu^{\rm 2+2B} \sim \sqrt{\lsnd} \,
\left(
\bav
\lambda &  \lambda & s_\eta   & c_\eta \\[0.2cm]
\cdot &  c_{\rm atm}^2 \, e^{i \beta} + s_{\rm atm}^2 \, e^{i \gamma} &
c_{\rm atm} s_{\rm atm} s_\eta (e^{i \beta} - e^{i \gamma}) &
c_{\rm atm}  s_{\rm atm}  c_\eta (e^{i \beta} - e^{i \gamma})
\\[0.2cm]
\cdot & \cdot & s_\eta^2 (c_{\rm atm}^2 \, e^{i \gamma} + s_{\rm atm}^2 \, e^{i \beta}) &
c_\eta s_\eta  (c_{\rm atm}^2 \, e^{i \gamma} + s_{\rm atm}^2 \, e^{i \beta}) \\[0.2cm]
\cdot & \cdot & \cdot &
c_\eta^2 (c_{\rm atm}^2 \, e^{i \gamma} + s_{\rm atm}^2 \, e^{i \beta})
\ea
\right) \\[0.3cm]
\simeq
\left(
\bav
\lambda &  \lambda & s_\eta   & c_\eta \\[0.2cm]
\cdot &  e^{i \beta} +  e^{i \gamma} &
s_\eta (e^{i \beta} - e^{i \gamma}) &
 c_\eta (e^{i \beta} - e^{i \gamma})
\\[0.2cm]
\cdot & \cdot & s_\eta^2 (e^{i \gamma} +e^{i \beta}) &
c_\eta s_\eta  (e^{i \gamma} + e^{i \beta}) \\[0.2cm]
\cdot & \cdot & \cdot &
c_\eta^2 (e^{i \gamma} + e^{i \beta})
\ea
\right)~,
\eea
where we set $\theta_{\rm atm}=\pi/4$. The difference with respect to
case (i) lies in the electron row of $m_\nu$.
Nevertheless, for $s_\eta \simeq c_\eta$ and $e^{i \beta} +  e^{i \gamma}$
one encounters again an approximate $L_e + L_\mu - L_\tau - L_s$ symmetry.
Setting $\eta=0$ gives
\be
m_\nu^{\rm 2+2B} \sim \sqrt{\lsnd} \,
\left(
\bav
\lambda & \lambda & 0 & 1 \\[0.2cm]
\cdot & e^{i \beta} +  e^{i \gamma} & 0 & e^{i \beta} - e^{i \gamma} \\[0.2cm]
\cdot & \cdot & 0 & 0 \\[0.2cm]
\cdot & \cdot & \cdot &  e^{i \beta} +  e^{i \gamma}
\ea
\right)~,
\ee
conserving $L_\tau$, if the $ee$ and $e \mu$ entries are not too strongly
suppressed. On the other hand, for $\eta=\pi/2$ one finds
\be
m_\nu^{\rm 2+2B} \sim \sqrt{\lsnd} \,
\left(
\bav
\lambda & \lambda & 1 & 0 \\[0.2cm]
\cdot & e^{i \beta} +  e^{i \gamma} & e^{i \beta} - e^{i \gamma} & 0 \\[0.2cm]
\cdot & \cdot & e^{i \beta} +  e^{i \gamma} & 0 \\[0.2cm]
\cdot & \cdot & \cdot &  0
\ea
\right)~,
\ee
i.e., a matrix conserving
$L_s$ when the $ee$ and $e \mu$ entries are not too strongly
suppressed.

\end{appendix}

\end{document}